\documentclass{article}

\usepackage{graphics}
\usepackage{units}
\usepackage{amsmath}
\usepackage{amssymb}
\usepackage{graphicx}
\usepackage{wasysym}
\usepackage{esint}
\usepackage{geometry}
\usepackage{color}
\usepackage[unicode=true,pdfusetitle,
 bookmarks=true,bookmarksnumbered=true,bookmarksopen=true,bookmarksopenlevel=1,
 breaklinks=false,pdfborder={0 0 0},backref=false,colorlinks=false]
 {hyperref}
\usepackage{breakurl}
\begin{document}
\title{Energetic optimization effects in single resonant tunneling  $GaAs$--nanoconverters}

\author{G. Valencia--Ortega\textsuperscript{a} and L. A. Arias--Hernandez\textsuperscript{b}\\
Departamento de F\'{i}sica, Escuela Superior de F\'{i}sica y Matem\'{a}ticas, Instituto\\
Polit\'{e}cnico Nacional, U. P. Zacatenco, edif. \#9, 2o Piso, Ciudad de M\'{e}xico,\\
07738, M\'{e}xico, gvalencia@esfm.ipn.mx; http://orcid.org/0000-0001-7453-5794\textsuperscript{a} and\\ larias@esfm.ipn.mx; 
http://orcid.org/0000-0003-4054-5446\textsuperscript{b}}
 
\maketitle

\begin{abstract}
Several models of thermionic energy nanoconverters have been proposed to study the transport phenomena that take place in electronic devices. For example, in resonant tunneling junctions those phenomena are manifested through the thermoelectric effects. The coupling between the electron flux and the heat flux in this type of semiconductor heterostructures, not only allows to obtain transport coefficients (electrical and thermal conductivities, and a Seebeck--like and Peltier--like coefficients), but also to study its operation as a thermionic generator or as a refrigerator within the context of irreversible thermodynamics. The existence of the characteristic steady states that can be reached by any linear energy converter led us to characterize a family of Seebeck--like coefficients, as well as establish bounds for the values of a kind of figure of merit $(Tz'_{D,I})$, both associated with the well-known operating regimes: minimum dissipation function, maximum power output, maximum efficiency and maximum compromise function. By taking as example an $Al_{x}GaAs/GaAs$ junction, we found that the transport coefficients depend strongly on temperature and the conduction band height, which can be modulated according to the selected operation mode.

\vspace{0.3cm}Keywords: non\textendash equilibrium thermodynamics, resonant tunneling devices, energetic optimization, figure of merit.

\end{abstract}

\section{Introduction}
\label{intro}
In nature there are many processes to convert one type of energy into another one. In most of them there is an amount of energy that is not usable (dissipated energy) to perform effective work. Within different contexts of Non-equilibrium Thermodynamics (NET) several authors have suggested ways to minimize the effects of this wasted energy within the systems \cite{Mahan98,SumithraSintes01}, for instance, through the operation modes that an energy converter can access \cite{AriasAresAngulo04,AriasPaezAngulo08}. The energy conversion processes that take place within the systems not only take notice its availability to transform one type of energy into another, as in the case of macroscopic thermal engines, but also to develop new thermodynamic applications in other type of devices \cite{Chanetal96,Mazumderetal98,Qu03,Katsiaetal09}. Currently, in the field of microelectronics, interest has been growing in studying the energetics in thermionic transport phenomena due to the charge carriers movement inside nanostructures \cite{Humphrey05,Luoetal13,AgarwalMuralidharan14,Zhouetal16,Yuetal16}.

The development of technology has led us to the miniaturization of some devices ranging from Brownian ratchet machines \cite{Humphreyetal02,Linkeetal02,Tu08} to nanoelectronic systems \cite{Goldhaber97,TurelLeeMaLikharev04,Tianetal07}. The rise of nanotechnology was mainly due to the necessity of processing information reliably, compactly, at high speed and at a low cost \cite{Sunetal98}. A proposal to reach these goals in the transmission of information is by optimizing the charge carriers dynamics in actual integrated circuits, for example, in semiconductor heterostructures known as resonant tunneling junctions \cite{TsuEsaki73,AverinLikharev91,WahoChenYamamoto96}. These systems have led to the study and development of the so-called Resonant Tunneling Diodes (RTDs) \cite{Sunetal98,ChangMendezTejedor91,IsmailMeyersonWang91,Slobodskyetal07}, which are used in some applications within digital circuits, for the purpose of establishing an optimization based on the design of these ones. However, the electron transport and their interaction with the unit cells, that make up these nanostructures, results in a heat flux, so RTDs could also be considered as energy converters.

\begin{figure}
\centering
\resizebox{0.65\textwidth}{!}{
\includegraphics[width=7cm]{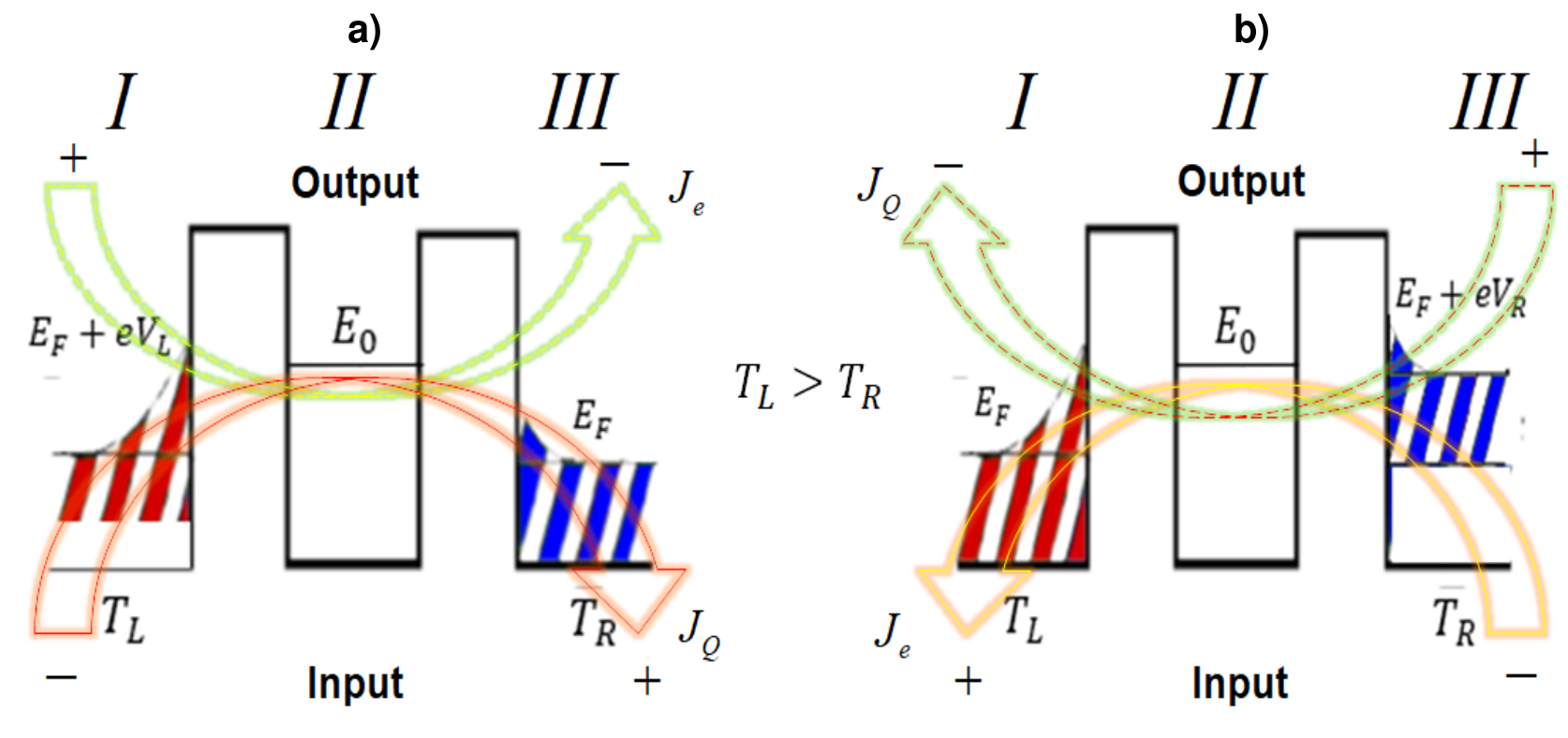}}
\caption{A Resonant Tunneling Diode seen as an $2\times2$ energy converter, according to its operation it can work as a generator (direct operation, where $J_{D2}$ is the driver flux and $J_{D1}$ is the driven one) or as a refrigerator, i.e, the flux $J_{I1}$ is now the  driver one, while the flux $J_{I2}$ is the driven one (inverse operation). In a), it is shown the basic scheme of an RTD whose operation is direct, in b) we show the inverse operation of the same RTD. $E_{F}$ is the Fermi level, $E_{0}$ is the single resonance energy level, $T_{L}$ and $T_{R}$ are the temperatures of the left and right reservoirs.}
\label{fig:rtdasconv}
\end{figure}

A Resonant Tunneling Diode can be understood as a system whose size becomes comparable with the electron wavelength, and consists of two barriers of finite potential among which is formed a potential well (see Fig. \ref{fig:rtdasconv}), with a single resonance energy value ($E_{0}\neq0$). This characteristic, as well as the multiple reflections, that the incident electrons can experience between the barriers' interfaces, produce the effect of resonance \cite{ChangEsakiTsu74,Bjorketal02}. In real nanostructures, the barriers are made of semiconductors with a large band gap but thin enough for the quantum tunnel effect to be achieved. In the scheme of Fig. \ref{fig:rtdasconv}, we consider that there are two reservoirs of electrons and energy on both the left and right sides of the RTD. Electron reservoirs are unbalanced with respect to each other by means of an effective unidirectional electric field. While energy reservoirs emulate a temperature gradient that promotes the heat flux due to the transport of electrons, so two coupled transport processes come out.

Within the context of Linear Irreversible Thermodynamics (LIT), these resonant tunneling devices, seen as thermodynamic systems, can be directly connected with the so-called linear energy converters \cite{AriasPaezAngulo08,CaplanEssig83,Stucki80,ValenciaArias17} which are understood as ``black boxes'' (see Fig. \ref{fig:rtdasconv}). The steady states that can be achived by the converter, are characterized through coupled fluxes (effects), and are classified as spontaneous or non-spontaneous and they are produced by external forces (causes). In particular, RTDs are considered as non--isothermal energy converters \cite{GonzalesArias19}, because of their analogy with the well-known thermoelectric phenomena. According to the operation objective of the device, the non--isothermal energy converters can be operated as direct (generator) or as inverse (refrigerator) \cite{Goupileetal11}.

In the linear regime, we can express the generalized fluxes $J_{i}$ $\left(i=D1, D2, I1, I2\right)$, following the works {by} Caplan and Essig \cite{CaplanEssig83}, in function of the generalized forces $X_{i}$ (Eq. \ref{eq:matrfenom}) to study the well-known cross-effect in this type of resonant tunneling devices, 
\begin{equation}
\left[\begin{array}{c}
J_{(D,I)1}\\
J_{(D,I)2}
\end{array}\right]=\left[\begin{array}{cc}
L_{11} & L_{12}\\
L_{21} & L_{22}
\end{array}\right]\left[\begin{array}{c}
X_{(D,I)1}\\
X_{(D,I)2}
\end{array}\right],\label{eq:matrfenom}
\end{equation}
the subscripts $D,I$ indicate whether the linear converter works directly (see Fig. \ref{fig:rtdasconv}a) or inversely (see Fig. \ref{fig:rtdasconv}b) and every $L_{ab}=\left(\nicefrac{\partial J_{(D,I)a}}{\partial X_{(D,I)b}}\right)_{eq}$ are the so-called Onsager phenomenological coefficients. When it is admitted the system given by Eq. \ref{eq:matrfenom} is symmetric ($L_{12}=L_{21}$), we can use two dimensionless parameters that simplify the study of the processes that occur in linear converters. The first one, called the coupling degree, measures the quality of coupling between spontaneous and non-spontaneous fluxes \cite{KedemCaplan65},
\begin{equation}
0\leq q^{2}=\frac{L_{12}^{2}}{L_{11}L_{22}}\leq1,\label{eq:couplparam}
\end{equation}
its {definition} comes directly from the second law of thermodynamics, that is, the entropy production $\sigma$ satisfies $\sigma=J_{D,I1}X_{D,I1}+J_{D,I2}X_{D,I2}\geq 0$. Then, in the linear regime (Eq. \ref{eq:matrfenom}) $\sigma$ turns out to be a positive semi-definite quadratic form. In the limit case $q\rightarrow0$, each flux is proportional to its proper conjugate force through its direct phenomenological coefficient, i.e, cross effects vanish and therefore the fluxes become independent. When $q\rightarrow1$, the fluxes tend to a fixed mechanistic stoichiometry relationship \cite{GonzalesArias19}.

The second parameter, known as the force ratio, was introduced by Stucki \cite{Stucki80} as a measure of the relationship between the forces ($X_{D1}$ or $X_{I2}$) associated with the driven fluxes and those, $X_{D2}$ or $X_{I1} $, associated with the driver fluxes. According to the operation of these non--isothermal converters, there are two definitions of this parameter \cite{GonzalesArias19}:
\begin{equation}
\begin{array}{c}
x_{D}=\sqrt{\frac{L_{11}}{L_{22}}}\frac{X_{D1}}{X_{D2}}\\ \\
x_{I}=\sqrt{\frac{L_{22}}{L_{11}}}\frac{X_{I2}}{X_{I1}}
\end{array},\label{eq:forcratiodirinv}
\end{equation}
where $x_{D,I}\in\left[-1,0\right]$. We also consider that in both the direct and inverse non--isothermal converter, the associated forces to driven and driver fluxes, correspond to a constant temperature gradient,
\begin{equation}
\begin{array}{r}
X_{D2}=\frac{1}{T_{R}}-\frac{1}{T_{L}}>0\\ \\
X_{I2}=\frac{1}{T_{L}}-\frac{1}{T_{R}}<0
\end{array},\label{eq:tempgraddirinv}
\end{equation}
with $T_{R}$ the temperature of the ``cold'' side and $T_{L}$ the temperature of the ``hot'' side. On the other hand, $X_{D1}$ and $X_{I1}$ are assumed as generalized forces that come from gradients of conservative potentials.

By using the parameters $q$ and $x_{D}$ for direct non--isothermal energy converters, the fluxes $J_{D1}$ and $J_{D2}$ are expressed as:
\begin{equation}
J_{D1}=\left(\frac{q+x_{D}}{q}\right)L_{12}X_{D2}=\zeta_{D1}(q,x_{D})L_{12}X_{D2}
\label{eq:fluxsdir}
\end{equation} and
\begin{equation}
J_{D2}=\left(1+qx_{D}\right)L_{22}X_{D2},
\label{eq:fluxdirb}
\end{equation}
similarly, with $q$ and $x_{I}$, the fluxes $J_{I1}$ and $J_{I2}$ for an inverse non--isothermal energy converter, they are,
\begin{equation}
J_{I1}=\left(\frac{qx_{I}+1}{qx_{I}}\right)L_{12}X_{I2}=\zeta_{I1}(q,x_{I})L_{12}X_{I2}
\label{eq:fluxsinv}
\end{equation} and
\begin{equation}
J_{I2}=\left(\frac{q+x_{I}}{x_{I}}\right)L_{22}X_{I2}.
\label{eq:fluxsinvb}
\end{equation}

Within the NET scheme, one of the main objectives is to establish commitments between the way to operate a linear energy converter and its design, in order to reduce the useless energy because of the non--ideal coupling between the fluxes. Those commitments can be studied through objective functions that come directly from the energetics of each linear converter. In this work (Section 2), we have considered an RTD (low-dimensional system) as non--isothermal linear converter. As a first approximation we have not taken into account magnetic effects in the transport of electrons. From the dynamics of electrons and considering the model of two--dimensional gas (2D), we wrote the phenomenological equations of the system, with the purpose of finding the analogous transport coefficients to those of the context of thermoelectricity (Seebeck coefficient, Peltier heat, etc.). Then, we established a connection between these transport coefficients and the phenomenological Onsager ones $L_{ab}$.

In recent years, within the context of nanoelectronics, other types of energy conversion models have been proposed and studied, for example the so-called ''Energy Selective Electron heat engines'' (ESE), which in practice are the most general version of RTDs. As several authors have pointed out, this class of thermionic devices has reached to its operating limits, bearing in mind transport mechanisms from a thermodynamic point of view. This topic of interest has been perfectly explained through the formalism of Finite Time Thermodynamics (FTT)  \cite{Luoetal13,Yuetal16,ZhouChenDingSun16}, which uses the energetics of the system (power output, efficiency and dissipated energy). We established additional objective functions (Section 3) which in turn are connected with characteristic operating regimes; in particular, the maximum compromise function regime.

In section 4, we used the information of the operation modes of non--isothermal linear energy converters, to find the optimization criteria that correspond to the characteristic functions in the context of LIT. The above, not only to establish a relationship between the optimal design and the way to operate the RTDs but also, to analyze the effects on energy transport when these devices are adapted to the required conditions by a specific steady state, with the aim of establishing guidelines for the choice of materials that allow RTDs to work with two different purposes (generator or refrigerator). In Section 5, we set bounds for a parameter (a kind of figure of merit) that is associated with both the design and operation of RTDs. Finally, in Section 6 we set forth the concluding remarks.

\section{Onsager equations for an RTD as a non-isothermal linear converter}
\label{sec:2}
All the formalisms that exist to study the semi-classical transport of particles are aimed at building the quantum transmission function \cite{Espsitoetal09,WangWuLu11,ChenDingSun11}. In this section, we focus on the Landauer--B\"{u}ttiker's model (LB) \cite{Datta95}, which assumes an elastic scattering of electrons, i.e, it is considered they do not lose energy during the collisions that experience due to semiconductor impurities. This approach has been reasonable for many nanostructures, since the wavelength of electrons is smaller than their mean free path in this type of materials. However, this assumption does not take into account dissipative effects such as phonon transport, magnetic interactions, etc.

The LB model also assumes that the electron energy spectrum in the $I$ and $III$ regions (see Fig. \ref{fig:rtdasconv}) satisfies the Fermi--Dirac equilibrium distribution, because the surroundings are considered large enough and the fluctuations are very small. Then, the current produced by the electron movement is defined in terms of their probability of transmission through the barriers. The distribution of Fermi--Dirac for each of the reservoirs is,
\begin{equation}
f_{L,R}(E)=\frac{1}{1+\exp\left[\nicefrac{E-\mu_{L,R}}{k_{B}T_{L,R}}\right]},\label{eq:FDdistr}
\end{equation}
where $k_{B}$ is the Boltzmann constant, $\mu_{L,R}=E_{F}+eV_{L,R}$ expresses the electrochemical potential of electrons and can take into account the energy due to voltage bias ($V_{L,R}$). The electron flux that appears from the $I$ region to the $III$ region can be written according to the elastic scattering approximation \cite{Datta95} as:
\begin{equation}
J_{e}=\frac{2e}{h}\sum_{n}\int M_{n}(E)\left[f_{p}(E)-f_{j}(E)\right]dE,\label{eq:curraproxLB}
\end{equation}
with $h$ the Planck's constant, $e$ the electron charge, the $2$ factor is due to the electron degeneration, $M_{n}(E)$ represents the function of the $nth$ transmission channel, added over the $n$ states that have the same energy and each $f_{j,p}(E)$ is given by Eq. \ref{eq:FDdistr}, where the subscript $p$ indicates that the density of charge carriers of the emitter ($L, R$) is greater than the density of the collector ($R, L$), both of them indicated with the subscript $j$.

On the other hand, the energy flux in form of heat due to the interactions of electrons with the metal ions, it is also considered to flow from the $I$ region to the $III$ region, assuming that $T_{L}>T_{R}$. Thus, the increase in energy because of the influence of the heat flux's direction (spontaneous or non-spontaneous) is expressed as $E-\mu_{L, R}$, then under the LB approximation, we have
\begin{equation}
J_{Q}=\frac{2e}{h}\sum_{n}\int M_{n}(E)\left(E-\mu_{L,R}\right)\left[f_{p}(E)-f_{j}(E)\right]dE.\label{eq:fluxheatapproxLB}
\end{equation}

The transmission function contains information about the scattered electrons in the materials that make up the different RTDs' nanostructures. It is common that in this type of devices there are resonance peaks. Therefore, we can approximate the transmission function by the Lorentzian function \cite{CoonBandaraZhao89}:
\begin{equation}
\sum_{n}M_{n}(E)\propto M(E)=\frac{\Gamma_{L}\Gamma_{R}}{\left(E-E_{0}\right)^{2}+\left(\frac{\Gamma_{L}+\Gamma_{R}}{2}\right)^{2}}.\label{eq:Lorentfunct}
\end{equation}
The $\Gamma_{L, R}$ parameters represent the leakage ratios of electron flux through the left or right barrier and, $E_{0}$ is the energy level of the single resonant state within the potential well.

\subsection{Transport equations of an RTD}
\label{sec:2.1}
Due to the structural properties of the materials that shape the contacts and the potential well of an RTD (LB approximation), we can assume the electrons movement as with that of a free particle with effective mass $m^{\ast}$. The energy of the sub-bands within the well are given by,
\begin{equation}
E=E_{z}+E_{\bot}=E_{z}+\frac{\hbar^{2}}{2m^{\ast}}\left(k_{x}^{2}+k_{y}^{2}\right),\label{eq:subbanener}
\end{equation}
we have considered that $Z$ is the direction of the electron propagation, perpendicular to the semiconductor layers (barriers), located in the plane $XY$, while $k_{x,y,z}$ are the components of the wave vector. In Eq. \ref{eq:curraproxLB} the sum over the transmission channels represents a density of states, which is considered $2$-dimensional,
\begin{equation}
\sum_{n}\rightarrow g_{2d}(E)=\frac{dn(E)}{dE}=\frac{Am^{\ast}}{2\pi\hbar^{2}}.\label{eq:2dstatdens}
\end{equation}

By substituting Eq. \ref{eq:2dstatdens} in Eq. \ref{eq:curraproxLB} and since the system is invariant in the perpendicular plane to the direction of propagation, we obtain:
\begin{equation}
J_{e}=\frac{eAm^{\ast}}{2\pi^{2}\hbar^{3}}\int_{eV_{L,R}}^{\infty}M(E_{z})dE_{z}\int_{0}^{\infty}\left[f_{p}\left(E_{\bot}+E_{z}\right)-f_{j}\left(E_{\bot}+E_{z}\right)\right]dE_{\bot},\label{eq:curraproxLB1}
\end{equation}
where $A$ is the face area of the barriers and $E_{\bot}=E_{x}+E_{y}$. In the $Z$ direction the accessible energies start from the lowest level of the band, which corresponds to $eV_{L,R}$.

With the contribution of the temperature gradient that is present in the transport of electrons along the RTD and represented by the temperatures of the reservoirs $T_{L}\neq T_{R}$, the mathematical expression that contains the distributions $f_{L}$ and $f_{R}$ in Eq. \ref{eq:curraproxLB1} rewrites the electric current as:
\begin{equation}
J_{e}=\frac{eAm^{\ast}}{2\pi^{2}\hbar^{3}}k_{B}\int_{eV_{L,R}}^{\infty}M(E_{z})\left\{ T_{p}\textrm{ln}\left[1+\exp\left(-\frac{E_{z}-\mu_{p}}{k_{B}T_{p}}\right)\right]-T_{j}\textrm{ln}\left[1+\exp\left(-\frac{E_{z}-\mu_{j}}{k_{B}T_{j}}\right)\right]\right\} dE_{z}.\label{eq:curraproxLB2}
\end{equation}
In order to obtain the generalized coefficients \cite{Onsager31}, we must linearize the Eq. \ref{eq:curraproxLB2}. Since the $\mu_{p}=E_{F}+eV_{p}$ and $\mu_{j}=E_{F}$, then $\nabla\mu=\mu_{p}-\mu_{j}=eV_{L,R}$ and therefore, $\nicefrac{\nabla\mu}{e}=V_{L,R}$. In addition, we define the temperatures of the left and right reservoirs as $T_{L}=T+\nicefrac{\Theta}{2}$ and $T_{R}=T-\nicefrac{\Theta}{2}$, where $\Theta$ is a small increase in the temperature of the system after moving it away from the equilibrium state whose temperature is $T$. Such that depending on the direction of heat flux: $\nabla T=T_{p}-T_{j}=\pm\Theta$. The $+$ sign corresponds to an RTD working as a generator, while the $-$ sign represents its operation as a refrigerator.

By rewriting the term in braces of the Eq. \ref{eq:curraproxLB2}, we obtain:
\begin{equation}
h(V_{L,R},\Theta)=T\ln\left(\frac{B}{C}\right)\pm\frac{\Theta}{2}\ln\left(BC\right),\label{eq:integT}
\end{equation}
with $B=B(V_{L,R},\Theta)=1+\exp\left(-\nicefrac{E_{z}-\mu_{p}}{k_{B}T_{p}}\right)$
and $C=C(\Theta)=1+\exp\left(-\nicefrac{E_{z}-\mu_{j}}{k_{B}T_{j}}\right)$. Now, if we take the function $h(V_{L,R},\Theta)$, we will be able to make a first-order Taylor series expansion in the vicinity of equilibrium state, and this effect will be directly reflected in $J_{e}$,
\begin{equation}
h(V_{L,R},\Theta)\approx h(0,0)+V_{L,R}\left(\frac{\partial h}{\partial V_{L,R}}\right)_{V_{L,R}=0,\Theta=0}+\Theta\left(\frac{\partial h}{\partial\Theta}\right)_{V_{L,R}=0,\Theta=0},\label{eq:linearapprox}
\end{equation}
the derivative with respect to the potential $V_{L,R}$, results,
\begin{equation}
\left(\frac{\partial h}{\partial V_{L,R}}\right)_{V_{L,R}=0,\Theta=0}=\frac{e}{k_{B}}f(E_{z}),\label{eq:deriv1}
\end{equation}
since $f_{p}(E_{z})=f_{j}(E_{z})=f(E_{z})$, as a consequence of $\mu_{L}=\mu_{R}$.

The derivative with respect to the temperature increase $\Theta$ is:
\begin{equation}
\begin{array}{c}
\left(\frac{\partial h}{\partial\Theta}\right)_{V_{L,R}=0,\Theta=0}=\pm\frac{T}{1+\exp\left(-\frac{E_{z}-E_{F}}{k_{B}T}\right)}\left[\frac{\left(E_{z}-E_{F}\right)\exp\left(-\frac{E_{z}-E_{F}}{k_{B}T}\right)}{k_{B}T^{2}}\right]\pm\frac{\ln(B^{2})}{2}\\ \\
=\pm\left\{ \frac{\left(E_{z}-E_{F}\right)}{k_{B}T}f(E_{z})+\ln\left[1+\exp\left(-\frac{E_{z}-E_{F}}{k_{B}T}\right)\right]\right\} 
\end{array},\label{eq:deriv2}
\end{equation}
owing to $\mu_{L}=\mu_{R}$, we have $B=C$. Hence, the electric current $J_{e}$ in the linear regime is rewritten,
\begin{equation}
\hspace{-0.5cm}J_{e}=\frac{eAm^{\ast}}{2\pi^{2}\hbar^{3}}k_{B}\int_{eV_{L,R}}^{\infty}M(E_{z})\left\{ V_{L,R}\left[\frac{e}{k_{B}}f(E_{z})\right]\pm\Theta\left[\frac{\left(E_{z}-E_{F}\right)}{k_{B}T}f(E_{z})+\ln\left[1+\exp\left(-\frac{E_{z}-E_{F}}{k_{B}T}\right)\right]\right]\right\} dE_{z}.\label{eq:currapproxlin}
\end{equation}

And the same thermoelectric--like transport coefficients appear \cite{Callen85}:
\begin{equation}
\begin{array}{c}
G=\frac{e^{2}Am^{\ast}}{2\pi^{2}\hbar^{3}}\int_{eV_{L,R}}^{\infty}f(E_{z})\left[\frac{\Gamma_{L}\Gamma_{R}}{\left(E_{z}-E_{0}\right)^{2}+\left(\frac{\Gamma_{L}+\Gamma_{R}}{2}\right)^{2}}\right]dE_{z}\\ \\
G\xi=\frac{eAm^{\ast}}{2\pi^{2}\hbar^{3}}k_{B}\int_{eV_{L,R}}^{\infty}\frac{\Gamma_{L}\Gamma_{R}}{\left(E_{z}-E_{0}\right)^{2}+\left(\frac{\Gamma_{L}+\Gamma_{R}}{2}\right)^{2}}\left\{ \frac{\left(E_{z}-E_{F}\right)}{k_{B}T}f(E_{z})+\ln\left[1+\exp\left(-\frac{E_{z}-E_{F}}{k_{B}T}\right)\right]\right\} dE_{z}
\end{array},\label{eq:transpcoeff1}
\end{equation}
where, the electrical conductivity $G$ and Seebeck--like (SL) coefficient generalized $\xi$ are expressed in terms of the Lorentzian propagation function. Under the same considerations, we substitute Eq. \ref{eq:2dstatdens} in Eq. \ref{eq:fluxheatapproxLB} and the heat flux $J_{Q}$ due to a difference in electrical potential along the RTD, is expressed as:
\begin{equation}
J_{Q}=\frac{Am^{\ast}}{2\pi^{2}\hbar^{3}}\int_{eV_{L,R}}^{\infty}M(E_{z})dE_{z}\int_{0}^{\infty}\left(E_{\bot}+E_{z}-\mu_{L,R}\right)\left[f_{p}\left(E_{\bot}+E_{z}\right)-f_{j}\left(E_{\bot}+E_{z}\right)\right]dE_{\bot},\label{eq:fluxheatapproxLB1}
\end{equation}
the integral that contains the $f_{L}$ and $f_{R}$ distributions in Eq. \ref{eq:fluxheatapproxLB1}, leads us to rewrite the heat flux,
\begin{equation}
\begin{array}{c}
J_{Q}=\frac{Am^{\ast}}{2\pi^{2}\hbar^{3}}k_{B}^{2}\int_{eV_{L,R}}^{\infty}M(E_{z})\left\{ T_{p}^{2}\left(\frac{E_{z}-\mu_{L,R}}{k_{B}T_{p}}\right)\ln\left[1+\exp\left(-\frac{E_{z}-\mu_{p}}{k_{B}T_{p}}\right)\right]-T_{p}^{2}\textrm{PolyLog}\left[2,-\exp\left(-\frac{E_{z}-\mu_{p}}{k_{B}T_{p}}\right)\right]-\right\} \\ \\
\left\{ T_{j}^{2}\left(\frac{E_{z}-\mu_{L,R}}{k_{B}T_{j}}\right)\ln\left[1+\exp\left(-\frac{E_{z}-\mu_{j}}{k_{B}T_{j}}\right)\right]+T_{j}^{2}\textrm{PolyLog}\left[2,-\exp\left(-\frac{E_{z}-\mu_{j}}{k_{B}T_{j}}\right)\right]\right\} dE_{z}
\end{array},\label{eq:fluxheatapproxLB2}
\end{equation}
with $\textrm{PolyLog}(2,-w)$ the polilogarithm function of second order. Again, we linearized the Eq. \ref{eq:fluxheatapproxLB2}, with the aim of obtaining the remaining transport coefficients and verify that Onsager's reciprocity relations are fulfilled. The term that appears between braces (Eq. \ref{eq:fluxheatapproxLB2}) can be defined as,
\begin{equation}
\begin{array}{c}
y\left(V_{L,R},\Theta\right)=\left(T\pm\frac{\Theta}{2}\right)^{2}\left\{ \frac{E_{z}-\mu_{L,R}}{k_{B}T_{p}}\ln\left(B\right)-\textrm{PolyLog}\left[2,-\left(B-1\right)\right]\right\} \\
\\
-\left(T\mp\frac{\Theta}{2}\right)^{2}\left\{ \frac{E_{z}-\mu_{L,R}}{k_{B}T_{j}}\ln\left(C\right)-\textrm{PolyLog}\left[2,-\left(C-1\right)\right]\right\} 
\end{array}, \label{eq:integV}
\end{equation}
with $B-1=\exp\left(-\nicefrac{E_{z}-\mu_{p}}{k_{B}T_{p}}\right)$, $C-1=\exp\left(-\nicefrac{E_{z}-\mu_{j}}{k_{B}T_{j}}\right)$. Similarly, we made a Taylor series expansion for $y\left(V_{L,R},\Theta\right)$ in the vicinity of the equilibrium state $\left(V_{L,R}=0,\Theta=0\right)$, whose effect will be reflected in the heat flux $J_{Q}$,
\begin{equation}
y(V_{L,R},\Theta)=y(0,0)+V_{L,R}\left(\frac{\partial y}{\partial V_{L,R}}\right)_{V_{L,R}=0,\Theta=0}+\Theta\left(\frac{\partial y}{\partial\Theta}\right)_{V_{L,R}=0,\Theta=0},\label{eq:linearapprox1}
\end{equation}
for this case, the derivative with respect to $V_{L,R}$ is:
\begin{equation}
\begin{array}{c}
\left(\frac{\partial y}{\partial V_{L,R}}\right)_{V_{L,R}=0,\Theta=0}=T^{2}\left\{ \frac{e\left(E_{z}-E_{F}\right)\exp\left(-\frac{E_{z}-E_{F}}{k_{B}T}\right)}{\left[1+\exp\left(-\frac{E_{z}-E_{F}}{k_{B}T}\right)\right]\left(k_{B}T\right)^{2}}+\frac{e\ln\left[1+\exp\left(-\frac{E_{z}-E_{F}}{k_{B}T}\right)\right]}{k_{B}T}\right\} \\ \\
=\frac{e\left(E_{z}-E_{F}\right)}{k_{B}^{2}}f(E_{z})+\frac{eT\ln\left[1+\exp\left(-\frac{E_{z}-E_{F}}{k_{B}T}\right)\right]}{k_{B}}
\end{array}.\label{eq:deriv3}
\end{equation}
Finally, the derivative with respect to $\Theta$ results,
\begin{equation}
\hspace{-0.5cm}\left(\frac{\partial y}{\partial\Theta}\right)_{V_{L,R}=0,\Theta=0}=\pm2\left\{ \frac{\left(E_{z}-E_{F}\right)\ln\left[1+\exp\left(-\frac{E_{z}-E_{F}}{k_{B}T}\right)\right]}{k_{B}}-T\textrm{PolyLog}\left[2,-\exp\left(-\frac{E_{z}-E_{F}}{k_{B}T}\right)\right]+\frac{\left(E_{z}-E_{F}\right)^{2}}{2k_{B}^{2}T}f(E_{z})\right\} .\label{eq:deriv4}
\end{equation}

The heat flux through the RTD $J_{Q}$, in the linear regime can be rewritten as,
\begin{equation}
\begin{array}{c}
J_{Q}=\frac{Am^{\ast}}{2\pi^{2}\hbar^{3}}k_{B}^{2}\int_{eV_{L,R}}^{\infty}M(E_{z})\left\{ eV_{L,R}\left[\frac{\left(E_{z}-E_{F}\right)}{k_{B}^{2}}f(E_{z})+\frac{T\ln\left[1+\exp\left(-\frac{E_{z}-E_{F}}{k_{B}T}\right)\right]}{k_{B}}\right]\pm2\Theta\left[\frac{\left(E_{z}-E_{F}\right)\ln\left[1+\exp\left(-\frac{E_{z}-E_{F}}{k_{B}T}\right)\right]}{k_{B}}-\right]\right\} \\ \\
\left\{ 2\Theta\left[T\textrm{PolyLog}\left[2,-\exp\left(-\frac{E_{z}-E_{F}}{k_{B}T}\right)\right]+\frac{\left(E_{z}-E_{F}\right)^{2}}{2k_{B}^{2}T}f(E_{z})\right]\right\} dE_{z}
\end{array},\label{eq:fluxheatapproxlin}
\end{equation}
the transport coefficients associated with heat flux can be identified as:
\begin{equation}
\hspace{-0.6cm}\begin{array}{c}
G\Pi=\frac{eAm^{\ast}}{2\pi^{2}\hbar^{3}}k_{B}^{2}\int_{eV_{L,R}}^{\infty}\frac{\Gamma_{L}\Gamma_{R}}{\left(E_{z}-E_{0}\right)^{2}+\left(\frac{\Gamma_{L}+\Gamma_{R}}{2}\right)^{2}}\left[\frac{\left(E_{z}-E_{F}\right)}{k_{B}^{2}}f(E_{z})+\frac{T\ln\left[1+\exp\left(-\frac{E_{z}-E_{F}}{k_{B}T}\right)\right]}{k_{B}}\right]dE_{z}\\ \\
\kappa_{V}=\frac{Am^{\ast}}{\pi^{2}\hbar^{3}}k_{B}^{2}\int_{eV_{L,R}}^{\infty}\frac{\Gamma_{L}\Gamma_{R}}{\left(E_{z}-E_{0}\right)^{2}+\left(\frac{\Gamma_{L}+\Gamma_{R}}{2}\right)^{2}}\left\{ \frac{\left(E_{z}-E_{F}\right)\ln\left[1+\exp\left(-\frac{E_{z}-E_{F}}{k_{B}T}\right)\right]}{k_{B}}-T\textrm{PolyLog}\left[2,-\exp\left(-\frac{E_{z}-E_{F}}{k_{B}T}\right)\right]+\frac{\left(E_{z}-E_{F}\right)^{2}}{2k_{B}^{2}T}f(E_{z})\right\} dE_{z}
\end{array},\label{eq:transpcoeff2}
\end{equation}
where a Peltier--like (PL) heat ($\Pi$) and the thermal conductivity measured at electric field equal to zero ($\kappa_V$) are also identified. They are written in terms of the Lorentzian propagation function.

From the Eqs. \ref{eq:transpcoeff1} and \ref{eq:transpcoeff2}, we can verify one relationship between the cross transport coefficients, in the same way as happens within the context of thermoelectricity, a second Thomson--like relation, $\Pi=T\xi$. This relation is a direct consequence of Onsager's reciprocity relationships ($L_{ab}=L_{ba}$) for phenomenological equations in the linear regime of an energy converter.
 
\subsection{Connection between transport coefficients and Onsager's phenomenological equations in an RTD}
\label{sec:2.2}
As well as thermoelectric phenomena (the Seebeck effect, the Peltier effect, the Thomson effect, the Joule effect and Ohm's law) are associated to a set of connected experiments to each other by the so-called Thomson relationships \cite{Callen85}, it is proposed by analogy, that the family of thermionic effects can be also related to the coupling of an electric current $J_{e}$ and a heat flux $J_{Q}$ in a given system, more specifically within a semiconductor. Due to spatial inhomogeneities, temperature and electrochemical potential gradients promote the appearance of the cross effects between existing fluxes; such that $J_{e}$ and $J_{Q}$ are written as:
\begin{equation}
\left[\begin{array}{c}
J_{e}\\
J_{Q}
\end{array}\right]=\left[\begin{array}{cc}
L_{11} & L_{12}\\
L_{21} & L_{22}
\end{array}\right]\left[\begin{array}{c}
\nicefrac{\nabla\mu}{eT}\\
\nabla\left(\nicefrac{1}{T}\right)
\end{array}\right].\label{eq:matrOnsag}
\end{equation}
This system shows Onsager's phenomenological equations (Eq. \ref{eq:matrfenom}) for a linear converter, where $J_{e}=J_{\left(D,I\right) 1}$ (electric current) and $J_{Q}=J_{\left(D,I\right) 2}$ (heat flux) are the generalized fluxes, while $X_{\left(D,I\right) 1}=\nicefrac{\nabla\mu}{eT}$ (the difference of electric potential) and $X_{\left(D,I\right) 2}=\nabla\left(\nicefrac{1}{T}\right)$ (the temperature gradient) are the forces that promote these fluxes. What can be measured experimentally in these systems does not correspond to the generalized equations, but the transport coefficients, then,
\begin{equation}
\left(\begin{array}{c}
J_{e}\\
J_{Q}
\end{array}\right)=\left(\begin{array}{cc}
G & G\xi\\
G\Pi & \kappa_{V}
\end{array}\right)\left(\begin{array}{c}
\nicefrac{\nabla\mu}{e}\\
\mp\nabla T
\end{array}\right).\label{eq:matrtrans}
\end{equation}

From the systems of Eqs. \ref{eq:matrOnsag} and \ref{eq:matrtrans}, it is possible to express the direct transport coefficients in terms of the phenomenological coefficients $L_{aa}$ as follows:
\begin{equation}
\begin{array}{c}
G=-\frac{J_{e}}{T\nabla X_{D,I1}}\mid_{\nabla T=0}=\frac{L_{11}}{T}\\ \\
\kappa_{V}=-\frac{J_{Q}}{\nabla T}\mid_{\nabla\mu=0}=\frac{L_{22}}{T^{2}}
\end{array},\label{eq:equiv1}
\end{equation}
so, if we match both systems of equations, the phenomenological coefficients are,
\begin{equation}
\begin{array}{c}
L_{11}=TG\\ \\
L_{12}=T^{2}G\xi\\ \\
L_{21}=TG\Pi\\ \\
L_{22}=T^{2}\kappa_{V}
\end{array}.\label{eq:equiv2}
\end{equation}

In the context of the LIT, the functions that describe the energetic evolution of thermodynamic systems can be expressed in terms of the dimensionless parameters $q$ and $x_{D,I}$, which at the same time are  functions of the transport coefficients. Therefore, the RTDs performance will be characterized by the above coefficients.

\section{Characteristic functions for generators and refrigerators}
\label{sec:3}
In order to study the energetic performance of various linear energy converters in different NET contexts, it has been found thermodynamic functions that describe the energetic evolution of these systems, considering the energy exchanges that they can experience with the surroundings, as well as the contribution due to internal couplings. In particular, the characteristic functions that emerge as a consequence of the study of non-isothermal linear converters show a dependence on the temperature ratio of the reservoirs ($\tau=\nicefrac{T_{R}}{T_{L}}$). For the two versions that have been characterized (generator or refrigerator), it is possible to choose combinations between these functions (dissipation function, the power output, efficiency) in order to obtain specific information about a system. These objective functions that correspond to operation regimes are the ecological function \cite{Angulo91}, the omega function \cite{Calvoetal01}, the efficient power \cite{Stucki80} and the Compromise Function \cite{AriasAngulo97,AnguloArias01,PartidoTornez06,AriasBarranco09,LevarioArias19}.

\subsection{Dissipation function}
\label{sec:3.1}
In the case of a non--isothermal energy converter, dissipation is defined as a function that not only measures the part of the energy that is used to achieve the coupling between the driven fluxes and the driver ones, but also contains information about the contribution of electron transport through the crystal lattice \cite{CaplanEssig83,GrootMazur62}.
\begin{equation}
\begin{array}{c}
\Phi_{D}=T_{R}\sigma_{D}=T_{R}\left(J_{D1}\frac{X_{D1}}{X_{D2}}+J_{D2}\right)X_{D2}=\eta_{rev}\left[x_{D}^{2}+2qx_{D}+1\right]L_{22}X_{D2}\\ \\
\Phi_{I}=T_{L}\sigma_{I}=T_{L}\left(J_{I1}\frac{X_{I1}}{X_{I2}}+J_{I2}\right)X_{I2}=-\frac{1}{COP_{rev}}\left[\frac{1+2qx_{I}+x_{I}^{2}}{x_{I}^{2}}\right]L_{22}X_{I2}
\end{array},\label{eq:dissfunctdirinv}
\end{equation}
with $\eta_{rev}=T_{C}X_{D2}=1-\tau$ the efficiency of a reversible thermal engine and $COP_{rev}=-T_{L}X_{I2}=\nicefrac{\tau}{1-\tau}$ the performance coefficient of a reversible refrigerator. From Eqs. \ref{eq:couplparam} and \ref{eq:forcratiodirinv}, the functions $\Phi_{D,I}$ as well as the rest of the energetic functions that will be described, are expressed in terms of $q$, $x_{D,I}$, $\eta_{rev}$ and $COP_{rev}$.

\subsection{Power output and Cooling power}
\label{sec:3.2}
Other characteristic functions that can be analyzed and exhibit the energy that a system can release as useful work, or can be used as input work to handle some process. Those functions are the power output and the cooling power, respectively,
\begin{equation}
\begin{array}{c}
P_{D}=-T_{R}J_{D1}X_{D1}=-\eta_{rev}x_{D}\left(x_{D}+q\right)L_{22}X_{D2}\\ \\
\dot{Q_{CI}}=J_{I2}=\left(1+\frac{q}{x_{I}}\right)L_{22}X_{I2}
\end{array}.\label{eq:powoutcoolpow}
\end{equation}
Cooling power is used to invert the heat flux, from the ``cold'' reservoir to the ``hot'' one. Thus, while the power output contributes negatively to the entropy production, the cooling power implies a positive contribution.

\subsection{Efficiency and coefficient of performance}
\label{sec:3.3}
The efficiency of a direct non-isothermal energy converter can be defined as $\eta\equiv\nicefrac{P_{D}}{J_{D2}}$, while the performance coefficient for a inverse energy converter is $COP\equiv\nicefrac{\dot{Q_{CI}}}{J_{I1}}$. Following the idea of Caplan and Essig \cite{CaplanEssig83}, both energetic performances can be expressed in terms of generalized fluxes and forces:
\begin{equation}
\begin{array}{c}
\eta_{D}=-\frac{T_{R}J_{D1}X_{D1}}{J_{D2}}=-T_{R}X_{D2}\frac{J_{D1}X_{D1}}{J_{D2}X_{D2}}=-\eta_{rev}\frac{x_{D}\left(x_{D}+q\right)}{1+qx_{D}}\\ \\
COP_{I}=\frac{J_{I2}}{T_{R}J_{I1}X_{I1}}=\left(\frac{1}{T_{R}X_{I2}}\right)\frac{J_{I2}X_{I2}}{J_{I1}X_{I1}}=-COP_{rev}\frac{x_{I}\left(x_{I}+q\right)}{1+qx_{I}}
\end{array}.\label{eq:efficcop}
\end{equation}
Although both Eqs. express the rate of converted energy and they have the same mathematical expression, efficiency is defined in the operating range $\left[0,1\right]$ and the coefficient
of performance in $\left[0,\infty\right)$.

\subsection{Compromise function}
\label{sec:3.4}
An alternative way to characterize thermodynamic processes in linear energy converters, it is through objective functions, which express different commitments between the system's processes variables. For example, there is an objective function that expresses a type of commitment between the power output and the dissipation or between cooling power and dissipation, known as Compromise Function (CF), in the context of energy converters \cite{AriasAngulo97,AnguloArias01,PartidoTornez06,AriasBarranco09,LevarioArias19},
\begin{equation}
\begin{array}{c}
C_{D}=\frac{P_{D}}{P_{D}^{MPO}}-\frac{T_{R}\sigma_{D}}{T_{R}\sigma_{D}^{MPO}}=\frac{-2x_{D}\left(x_{D}+q\right)}{q^2}-\frac{4\left(x_{D}^2+2qx_{D}+1\right)}{4-3q^2}\\ \\
C_{I}=\frac{\dot{Q_{CI}}}{\dot{Q_{CI}^{*}}}-\frac{T_{L}\sigma_{I}}{T_{L}\sigma_{I}^{*}}=\frac{x_{I}^{*}\left(x_{I}+q\right)}{x_{I}\left(x_{I}^{*}+q\right)}-\frac{\left(x_{I}^{*}\right)^2\left(x_{I}^{2}+2qx_{I}+1\right)}{x_{I}^{2}\left(\left(x_{I}^{*}\right)^2+2qx_{I}^{*}+1\right)}
\end{array},\label{compfunct}
\end{equation}
where $x_{I}^{*}$ comes from solving the constrain $COP=\left(\nicefrac{1}{2}\right)COP_{MCOP}$. If a linear converter operates beneath this criterion, we would obtain that power output or cooling power not decrease too much, and the dissipated energy not reach large values.

\begin{figure}
\centering
\resizebox{0.7\textwidth}{!}{
\includegraphics[width=6cm]{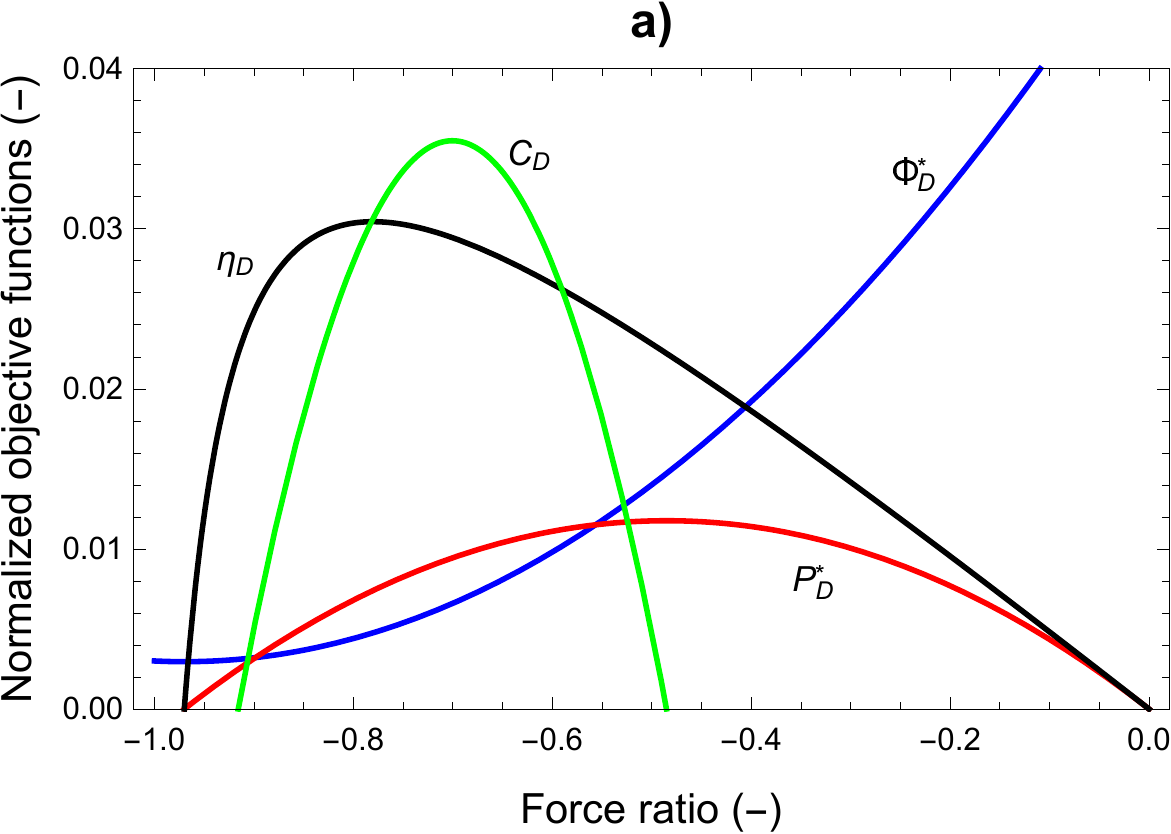}\includegraphics[width=6cm]{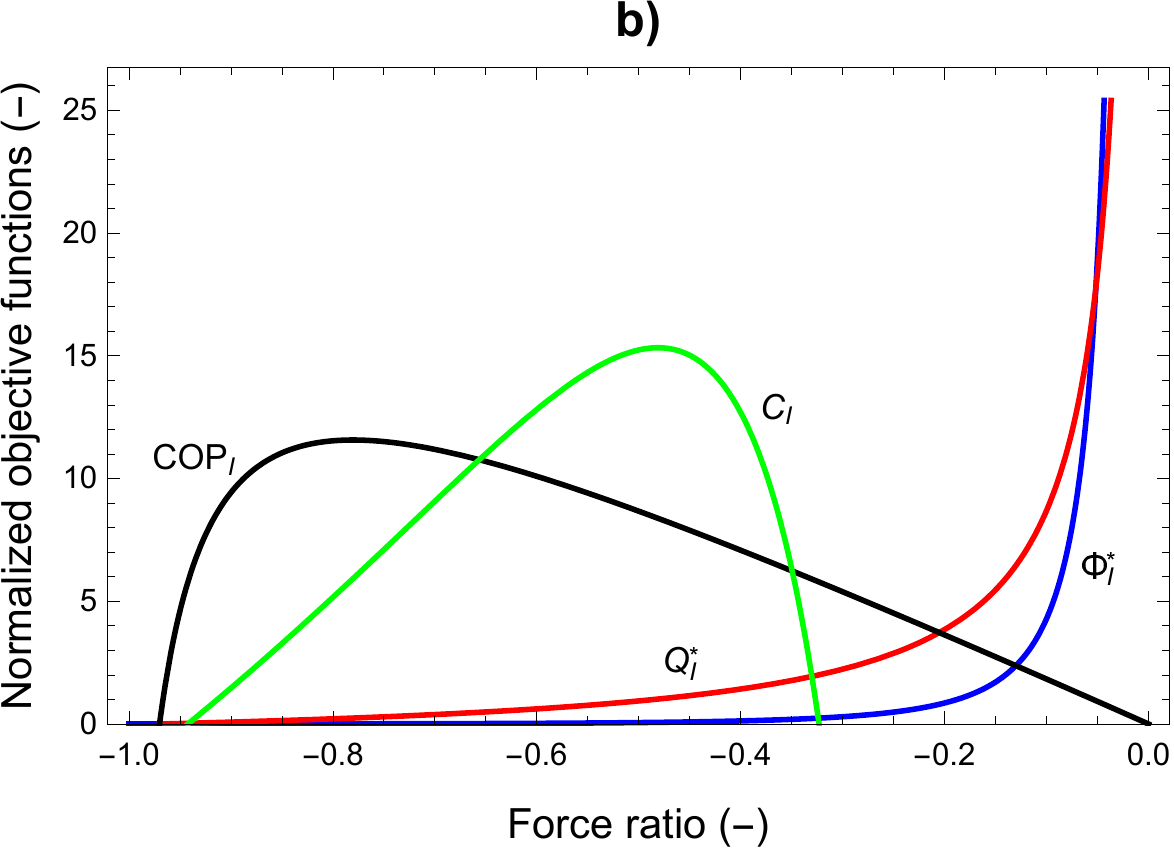}}
\caption{Characteristic functions normalized by the fixed value $L_{22}X_{\left(D,I\right)2}$ plotted against the force ratio $x_{D,I}$. a) For direct linear energy converters: dissipation function $\Phi_{D}^{*}$, power output $P_{D}^{*}$, efficiency $\eta_{D}$ (not normalized) and compromise function $C_{D}$ (not normalized but multiplied by a scaling factor of 0.1). b) For  inverse linear energy converters: dissipation function $\Phi_{I}^{*}$, cooling power $\dot{Q}_{CI}^{*}$, coefficient of performance $COP_{I}$ (not normalized) and compromise function $C_{I}$ (not normalized), all these functions are sketched for $q=0.97$ and $\tau=0.94$.}
\label{fig:charfunctdirinv}
\end{figure}

The behavior of some of the previous characteristic functions is monotonous decreasing ($\Phi_{D,I}$, $\dot{Q_{CI}}$), but mostly they are convex functions with a well defined maximum as is shown in Fig. \ref{fig:charfunctdirinv}, for both of the versions of non--isothermal linear energy converters. This extreme behavior of the characteristic functions suggests optimal operation modes that correspond to different working regimes.
 
With the help of the above mentioned characteristic functions, which arise from the fundamental equations (first and second laws of thermodynamics), we made the connection under the same hypotheses that are valid for macroscopic systems with low-dimensional systems, which use semi--classical and linear physical models. To establish the trade--offs between ways to operate the RTDs with its construction via the phenomenological coefficients of Onsager.

\section{Energetic optimization of an RTD under different operation regimes}
\label{sec:4}
From the graphs in Fig. \ref{fig:charfunctdirinv}, we can observe that almost all the characteristic functions reach maximum or minimum values. So, if $F_{D,I}(q,x_{D,I})$ denotes such energetic functions, the expression $\nicefrac{dF_{D,I}}{dx_{D,I}}\mid_{x_{D,I}^{\text{\#}}}=0$ accounts for the value of force ratio that maximizes or minimizes a particular function. In Tab. \ref{tab:optvalforcrat} different operation modes are shown.
\begin{table}
\begin{centering}
\caption{Optimum values for the force ratio of some operating regimes: minimum dissipation function ($mdf$), maximum power output ($MPO$), maximum efficiency ($M\eta$) and maximum compromise function ($MC$).\vspace*{0.25cm}}
\label{tab:optvalforcrat}
\begin{tabular}{|c|c|}
\hline 
Direct linear energy converter & Inverse linear energy converter\tabularnewline
\hline 
\hline 
$x_{D}^{mdf}=-q$ & $x_{I}^{mdf}=-\frac{1}{q}$\tabularnewline
\hline 
$x_{D}^{MPO}=-\frac{q}{2}$ & \textendash{}\tabularnewline
\hline 
$x_{D}^{M\eta}=-\frac{q}{1+\sqrt{1-q^{2}}}$ & $x_{I}^{MCOP}=-\frac{q}{1+\sqrt{1-q^{2}}}$\tabularnewline
\hline 
$x_{D}^{MC}=-\frac{q\left(q^{2}-4\right)}{4\left(q^{2}-2\right)}$ & $x_{I}^{MC}=-\frac{2q}{4\left(1+\sqrt{1-q^{2}}\right)-q^{2}}$\tabularnewline
\hline 
\end{tabular}
\par\end{centering}
\end{table}
It can be noted the relation between the operation of a linear converter expressed by a particular $x_{D,I}^{\#}$ and its construction (design) given by $q(L_{ab})$. This trade--off is well known for thermodynamic systems whose main objective is the transformation of an energy input into a useful  energy output. From Eqs. \ref{eq:fluxsdir} and \ref{eq:fluxsinv} that describe the generalized fluxes, we can notice they depend of the operation mode by means of $x_{D,I}$. In particular, if we take $J_{e}=J_{\left(D,I\right)1}(q,x_{D,I}^{\#})=\zeta_{D,I}(q,x_{D,I}^{\#})L_{12}\nabla(\nicefrac{1}{T})$, then there is a family of SL coefficients that correspond to different operating regimes, i.e, the energy input transferred through the barriers (contacts) in the form of work (voltage) to its surroundings, with a fixed value $J_{e}$ corresponding to a particular steady state.

The SL coefficient for the general case $J_{\left(D,I\right)1}=J_{\left(D,I\right)1}(q,x_{D,I}^{\#})$ is defined:
\begin{equation}
\xi_{D,I}^{\#}\equiv\pm\left.\frac{\partial V_{L,R}}{\partial T}\right|_{J_{\left(D,I\right)1}=cte}=\mp\left[\zeta_{D,I}(q,x_{D,I}^{\#})-1\right]\frac{L_{12}}{TL_{11}},\label{eq:coefseeb}
\end{equation}
where upper sign corresponds to the case of thermionic generator and lower sign is for thermionic refrigerator. The PL effect is understood as the heat flux promoted only by the existence of a fixed electric current in isothermal conditions: $J_{\left(D,I\right)1}=J_{\left(D,I\right)1}(q,x_{D,I}^{\#})$,
\begin{equation}
\Pi_{D,I}\equiv\left.\frac{J_{Q}}{J_{e}}\right|_{\nabla T=0}=\frac{L_{21}}{L_{11}}.\label{eq:coefpelt}
\end{equation}

\subsection{Thomson's second relation for an RTD as generator}
\label{sec:4.1}
When it is desired to operate an RTD as a direct energy converter, the behavior of the transport coefficients ($\xi,\Pi$) can be studied under any optimal accessible operation mode, corresponding to each of the different $x_{D}^{\#}$ of the Tab. \ref{tab:optvalforcrat}. From Eqs. \ref{eq:transpcoeff1} and \ref{eq:coefseeb} the SL coefficient can be characterized for a particular current $J_{e}^{\#}$,
\begin{equation}
\begin{array}{c}
\xi_{D}^{\#}=\frac{G\xi^{\#}}{G}=-\frac{k_{B}}{e}\left(\frac{\int_{eV_{L}}^{\infty}\frac{\Gamma_{L}\Gamma_{R}}{\left(E_{z}-E_{0}\right)^{2}+\left(\frac{\Gamma_{L}+\Gamma_{R}}{2}\right)^{2}}\left\{ \frac{\left(E_{z}-E_{F}\right)}{k_{B}T}f(E_{z})+\ln\left[1+\exp\left(-\frac{E_{z}-E_{F}}{k_{B}T}\right)\right]\right\} dE_{z}}{\int_{eV_{L}}^{\infty}f(E_{z})\left[\frac{\Gamma_{L}\Gamma_{R}}{\left(E_{z}-E_{0}\right)^{2}+\left(\frac{\Gamma_{L}+\Gamma_{R}}{2}\right)^{2}}\right]dE_{z}}\right)\left[\zeta_{D}(q,x_{D}^{\#})-1\right]\\ \\
=-\frac{k_{B}}{e}\left(\frac{\int_{eV_{L}}^{\infty}\frac{\left(\frac{\Gamma}{2}\right)^{2}}{\left(E_{z}-E_{0}\right)^{2}+\left(\frac{\Gamma}{2}\right)^{2}}\left\{ \frac{\left(E_{z}-E_{F}\right)}{k_{B}T}f(E_{z})+\ln\left[1+\exp\left(-\frac{E_{z}-E_{F}}{k_{B}T}\right)\right]\right\} dE_{z}}{\int_{eV_{L}}^{\infty}f(E_{z})\left[\frac{\left(\frac{\Gamma}{2}\right)^{2}}{\left(E_{z}-E_{0}\right)^{2}+\left(\frac{\Gamma}{2}\right)^{2}}\right]dE_{z}}\right)\left[\zeta_{D}(q,x_{D}^{\#})-1\right]
\end{array}.\label{eq:coefseebdir}
\end{equation}

\begin{figure}
\centering
\resizebox{0.7\textwidth}{!}{
\includegraphics[width=7cm]{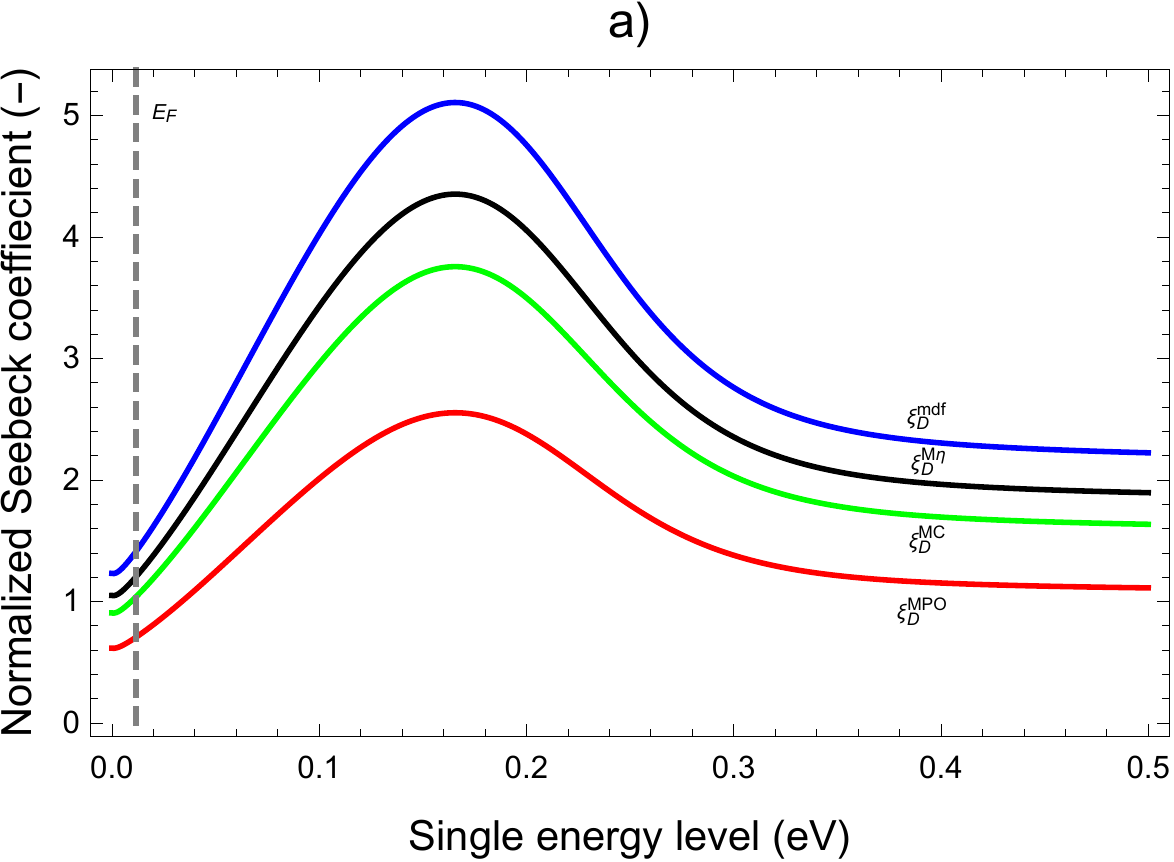}\includegraphics[width=7cm]{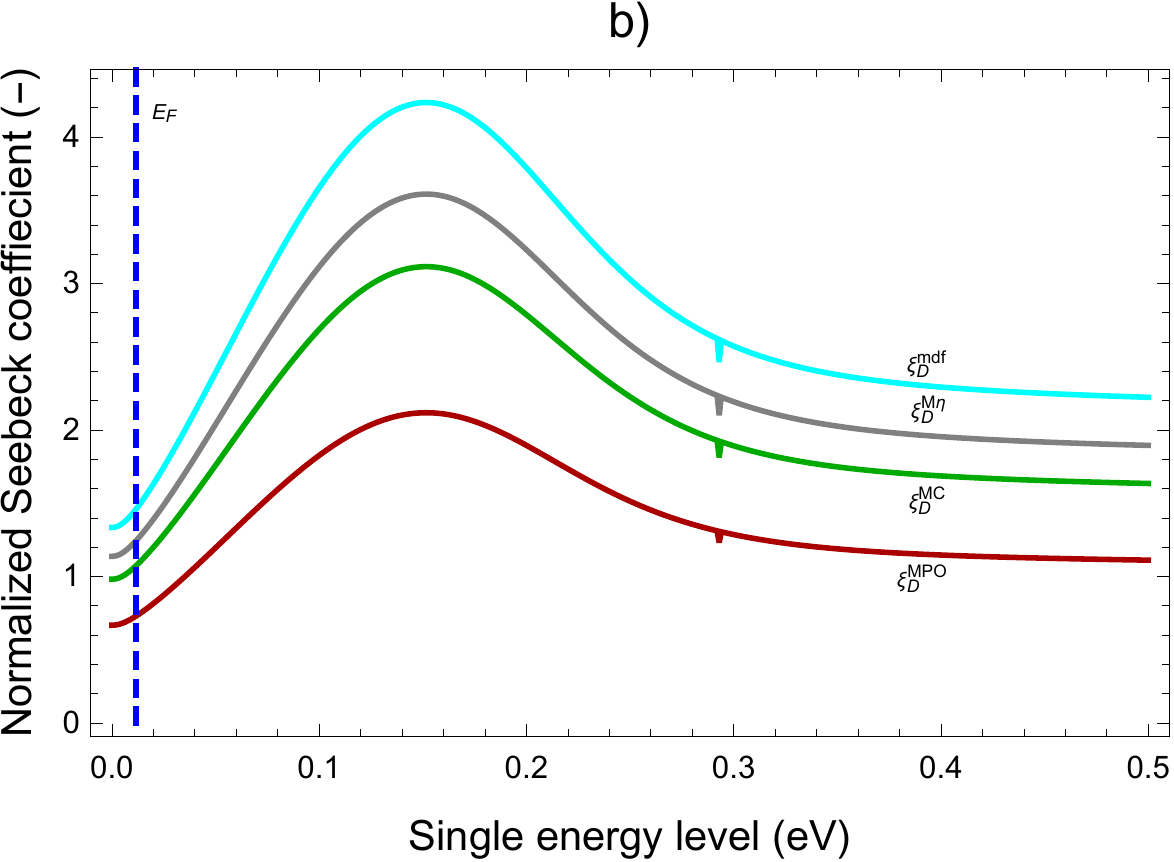}\par}
\centering
\resizebox{0.7\textwidth}{!}{
\includegraphics[width=7cm]{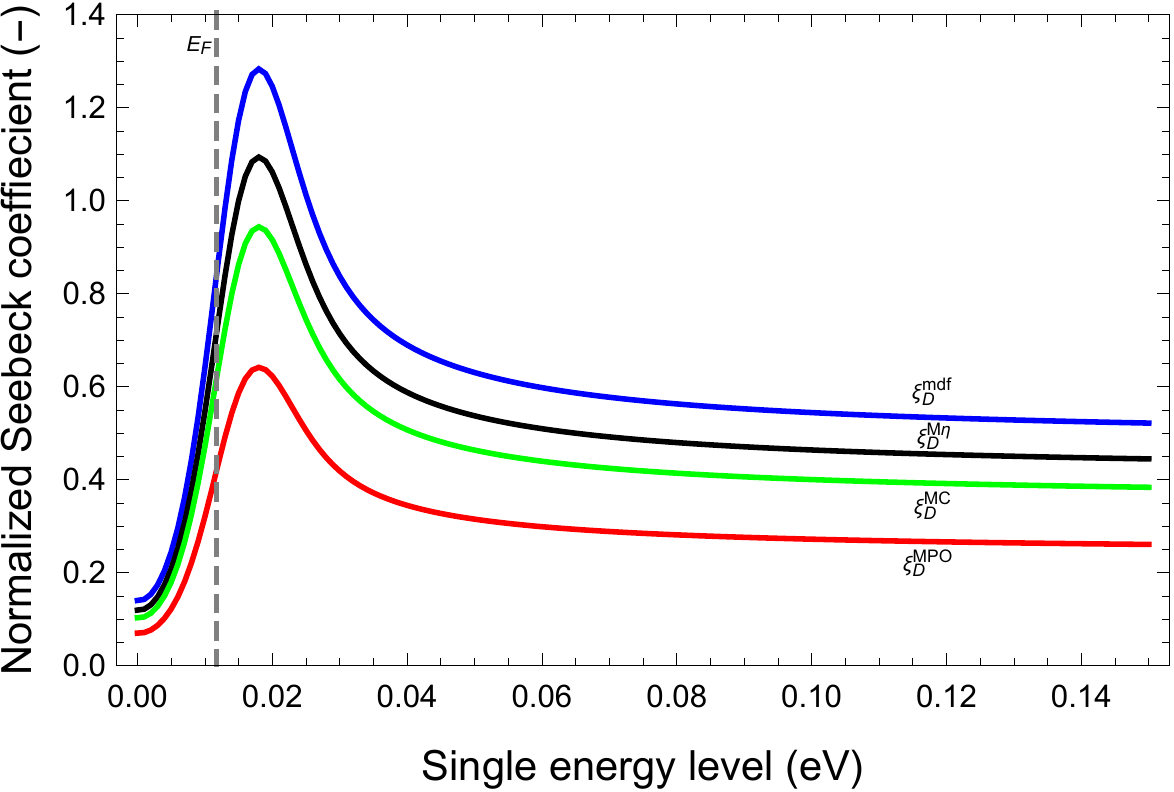}\includegraphics[width=7cm]{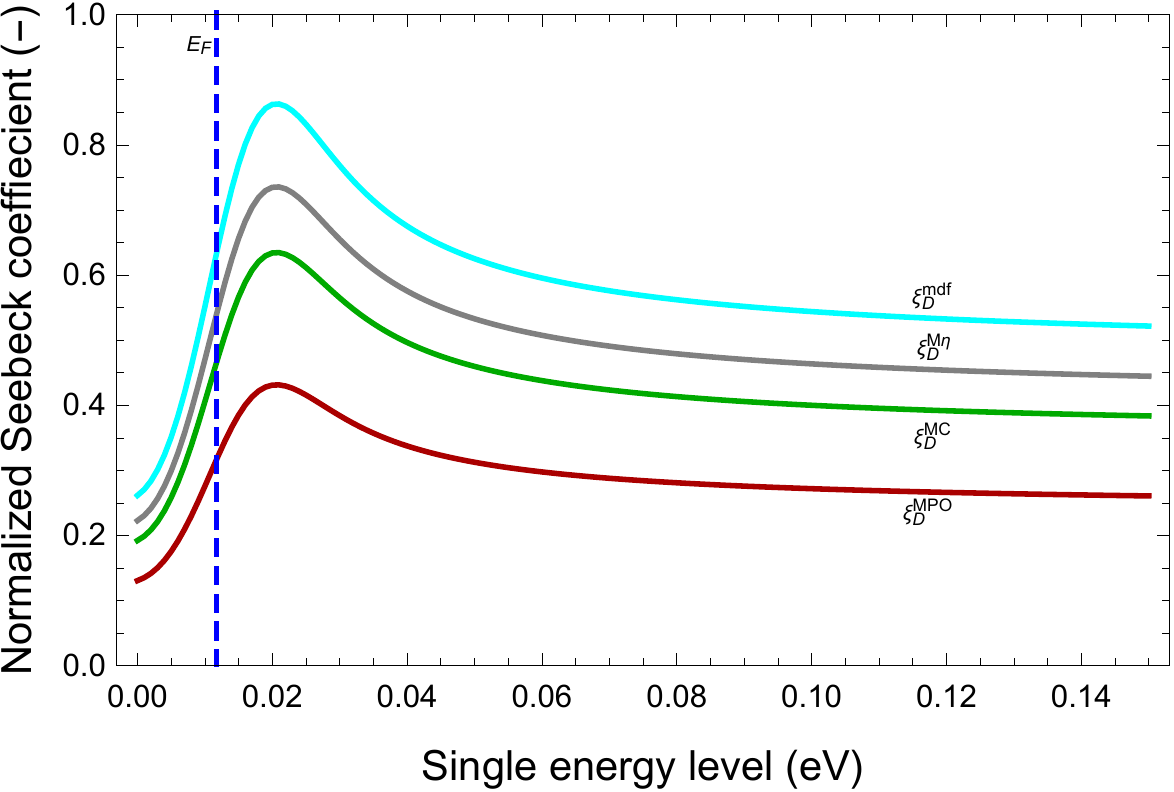}\par}
\centering
\resizebox{0.7\textwidth}{!}{
\includegraphics[width=7cm]{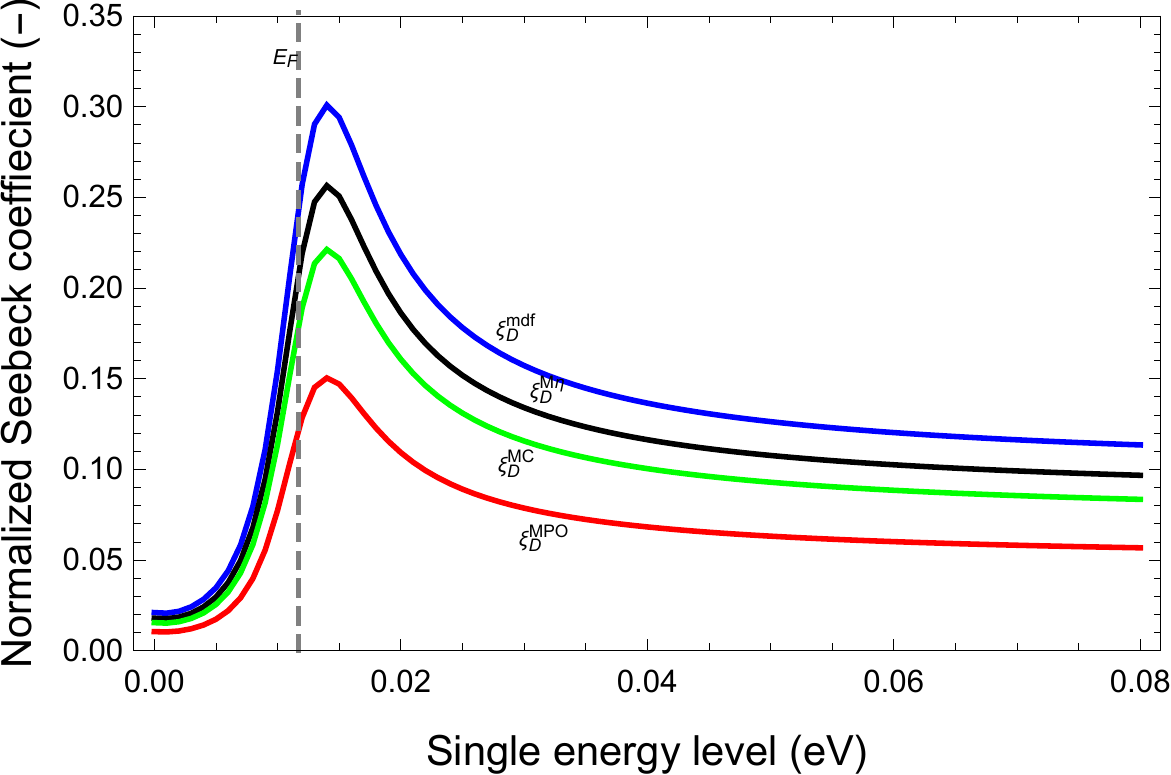}\includegraphics[width=7cm]{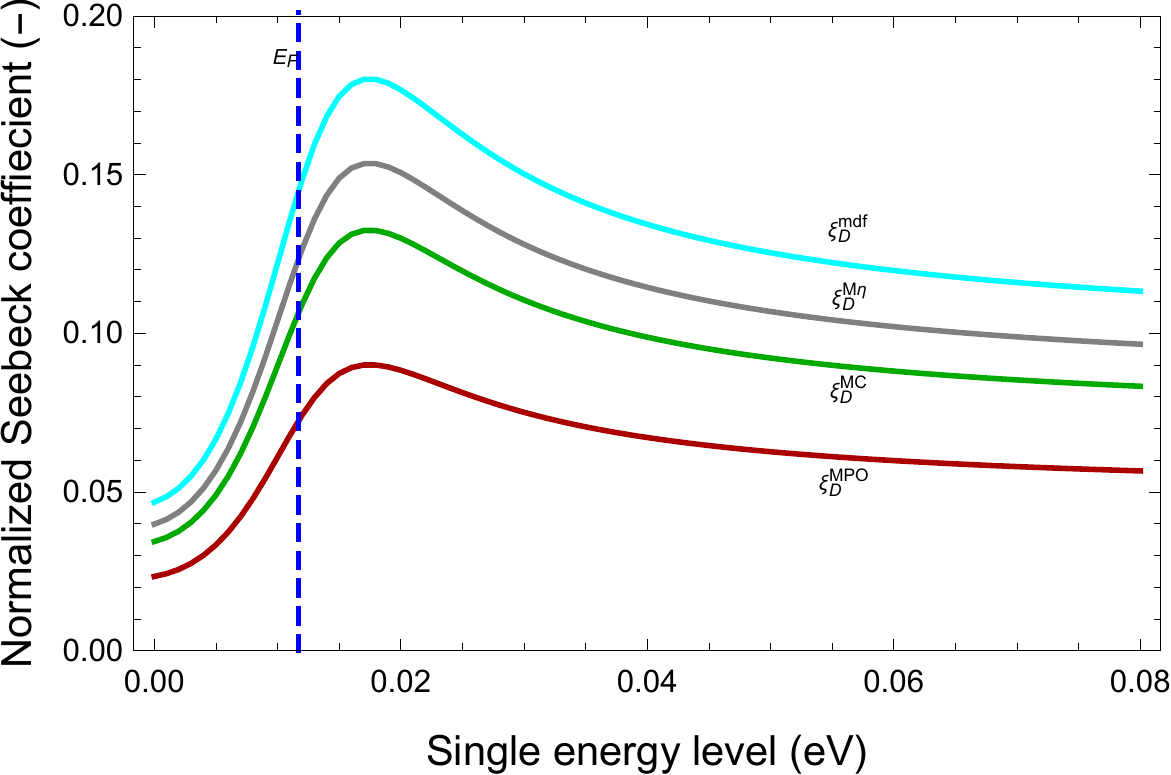}\par}
\caption{Graphs of the normalized SL coefficient family $\nicefrac{\xi_{D}^{\#}}{\left(\frac{k_{B}}{e}\right)}$ evaluated for each defined work regime for a direct linear energy converter ($mdf$, $MPO$, $M\eta$, $MC$) plotted as a function of the single resonance energy position of the well $E_{0}$. In a), the case of symmetric barriers is considered, $\Gamma=7\,\textrm{mV}$, $q=0.98$ and temperature values at $300\,\textrm{K}$, $20\,\textrm{K}$ and $4\,\textrm{K}$ in descending order. In b), the case of asymmetric barriers is shown, $\Gamma_{L}=7.5\,\textrm{mV}$, $\Gamma_{R}=8.5\,\textrm{mV}$,
$q=0.98$ for the same values of temperatures.}
\label{fig:famseebcoef}
\end{figure}

We have considered that the barriers that delimit the regions $I-II$ and $II-III$ (see Fig. \ref{fig:rtdasconv}a) have two configurations: the same width and therefore the same tunneling ratio ($\Gamma_{L}=\Gamma_{R}=\nicefrac{\Gamma}{2}$) or two different widths, that is, different tunneling ratios ($\Gamma_{L}\neq\Gamma_{R}$). In addition, for a RTD, the SL coefficient does not depend on the reduced mass, nor on the plate's area. For steady states compatible with the known operating regimes, the following family of SL coefficients can be defined,
\begin{equation}
\begin{array}{c}
\xi_{D}^{mdf}=\xi_{D}\\ \\
\xi_{D}^{MPO}=\frac{1}{2}\xi_{D}\\ \\
\xi_{D}^{M\eta}=\frac{1}{1+\sqrt{1-q^{2}}}\xi_{D}\\ \\
\xi_{D}^{MC}=\frac{\left(q^{2}-4\right)}{4\left(q^{2}-2\right)}\xi_{D}
\end{array},\label{eq:optcoefseebdir}
\end{equation}
where $\xi_{D}$ is the SL coefficient measured at $J_{e}=0$. In the curves of Fig. \ref{fig:famseebcoef} it is shown on the one hand that the family of generated SL coefficients reaches a maximum for the same value of the single resonance energy $E_{0}$, and on the other hand for $E_ {0}>E_ {F}$ or $E_{0}<E_{F}$ the value of SL coefficients is positive, this means there is always an effective electron current from the emitter at temperature $T_{L}$ towards the collector at temperature $T_{R}$. Because of the Fermi energy ($E_{F}$) is kept at a fixed value, the variation of the single resonance energy level $E_ {0}$, that is shown in Fig. \ref{fig:famseebcoef}, is reflected as a change in the conduction band width (barrier height) of the semiconductor. In addition, it is observed that the behavior of the resonance energy with respect to temperature satisfies the condition $E_{0}\approx E_{F}$ if $T\rightarrow0$.

According to the experimental values reported in \cite{ChangMendezTejedor91}, for a heterostructure consisting of two barriers of $Al_{x}Ga_{1-x}As$ separated by a well of $GaAs$ with a fixed width of $7\,\textrm{nm}$ and embedded between two doped electrodes of $n-GaAs$, the constant value of the well width leads us to a fixed value for $E_{F}=11.7\,\textrm{meV}$. In addition, to ensure non-negative values of electrical conductivity due to electronic interactions within the well, the $\Gamma$ parameter must be of the following order $7-9\,\textrm{mV}$.

On the other hand, since $L_{12}=L_{21}$ is implicitly fulfilled in the transport equations (Eqs. \ref{eq:currapproxlin} and \ref{eq:fluxheatapproxlin}), a family of Thomson second relations is generated for each operating regime,

\begin{equation}
\begin{array}{c}
\Pi_{D}^{mdf}=T\xi_{D}\\ \\
\Pi_{D}^{MPO}=2T\xi_{D}\\ \\
\Pi_{D}^{M\eta}=\left(1+\sqrt{1-q^{2}}\right)T\xi_{D}\\ \\
\Pi_{D}^{MC}=\frac{4\left(q^{2}-2\right)}{\left(q^{2}-4\right)}T\xi_{D}
\end{array},\label{eq:optcoefpelt}
\end{equation}

\subsection{Thomson's second relation for an RTD as refrigerator}
\label{sec:4.2}
On the other hand, if an RTD is operated as an inverse energy converter, the characterization of the transport coefficients ($\xi,\Pi$) presents a behavior similar to the direct one, with only a difference, there are fewer accessible operating regimes, they are associated with each $x_{I}^{\#}$ (see Tab. \ref{tab:optvalforcrat}). Then, from the same Eqs. \ref{eq:transpcoeff1} and \ref{eq:coefseeb}, the Seebeck coefficient for a current $J_{e}^{\#}$ is:
\begin{equation}
\xi_{I}^{\#}=\frac{k_{B}}{e}\left(\frac{\int_{eV_{R}}^{\infty}\frac{\Gamma_{L}\Gamma_{R}}{\left(E_{z}-E_{0}\right)^{2}+\left(\frac{\Gamma_{L}+\Gamma_{R}}{2}\right)^{2}}\left\{ \frac{\left(E_{z}-E_{F}\right)}{k_{B}T}f(E_{z})+\ln\left[1+\exp\left(-\frac{E_{z}-E_{F}}{k_{B}T}\right)\right]\right\} dE_{z}}{\int_{eV_{R}}^{\infty}f(E_{z})\left[\frac{\Gamma_{L}\Gamma_{R}}{\left(E_{z}-E_{0}\right)^{2}+\left(\frac{\Gamma_{L}+\Gamma_{R}}{2}\right)^{2}}\right]dE_{z}}\right)\left[\zeta_{I}(q,x_{I}^{\#})-1\right].\label{eq:coefseebinv}
\end{equation}

If a linear energy converter operates inversely, the flux of electrons is able to extract energy from the collector ($I$) to the emitter ($II$) (see Fig. \ref{fig:rtdasconv}b). Again, the SL coefficient results be independent of device geometrical properties. We took the same two cases: $\Gamma_{L}=\Gamma_{R}=\nicefrac{\Gamma}{2}$ and $\Gamma_{L}\neq\Gamma_{R}$. The steady states compatible with the three  operating regimes ($mfd$, $MCOP$ and $MC$) define another family of SL coefficients,

\begin{equation}
\begin{array}{c}
\xi_{I}^{mfd}=-\xi_{I}\\ \\
\xi_{I}^{MCOP}=-\frac{1+\sqrt{1-q^{2}}}{q^{2}}\xi_{I}\\ \\
\xi_{I}^{MC}=-\left[\frac{4\left(1+\sqrt{1-q^{2}}\right)-q^{2}}{2q^{2}}\right]\xi_{I}
\end{array}.\label{eq:optcoefseebinv}
\end{equation}

\begin{figure}
\centering
\resizebox{0.7\textwidth}{!}{
\includegraphics[width=7cm]{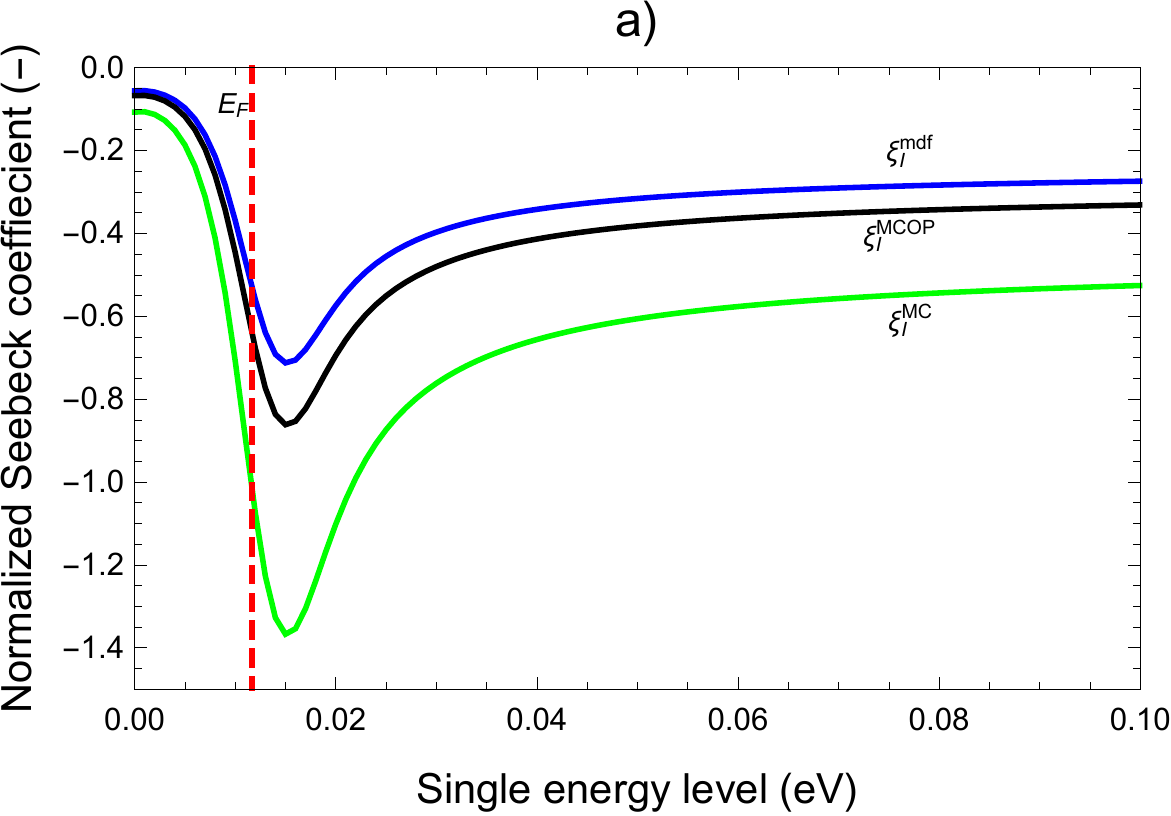}\includegraphics[width=7cm]{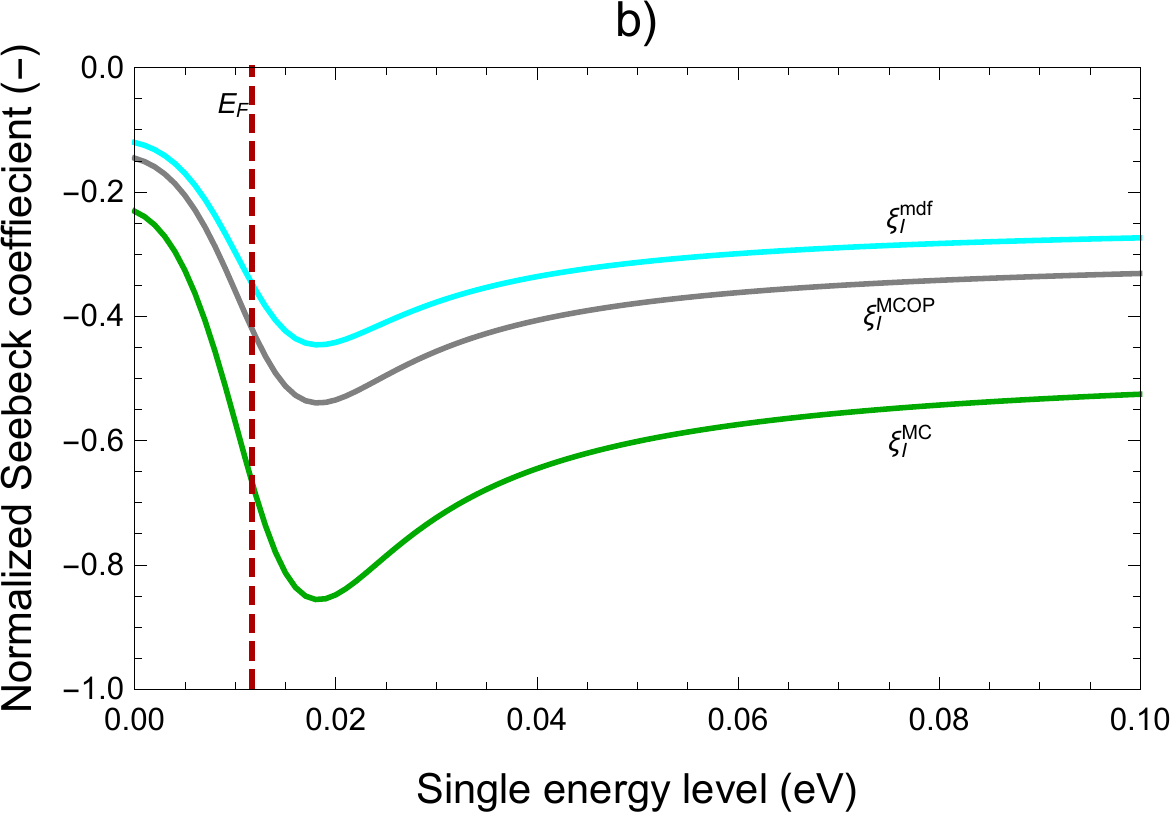}\par}
\centering
\resizebox{0.7\textwidth}{!}{
\includegraphics[width=7cm]{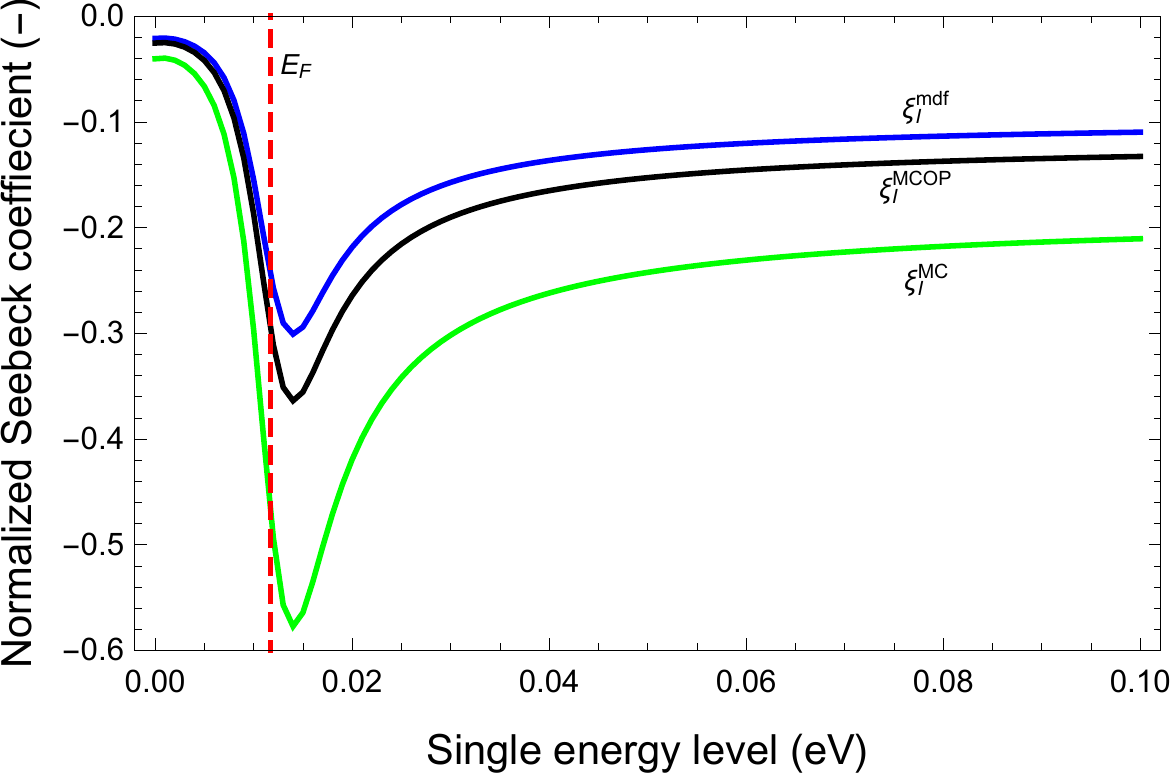}\includegraphics[width=7cm]{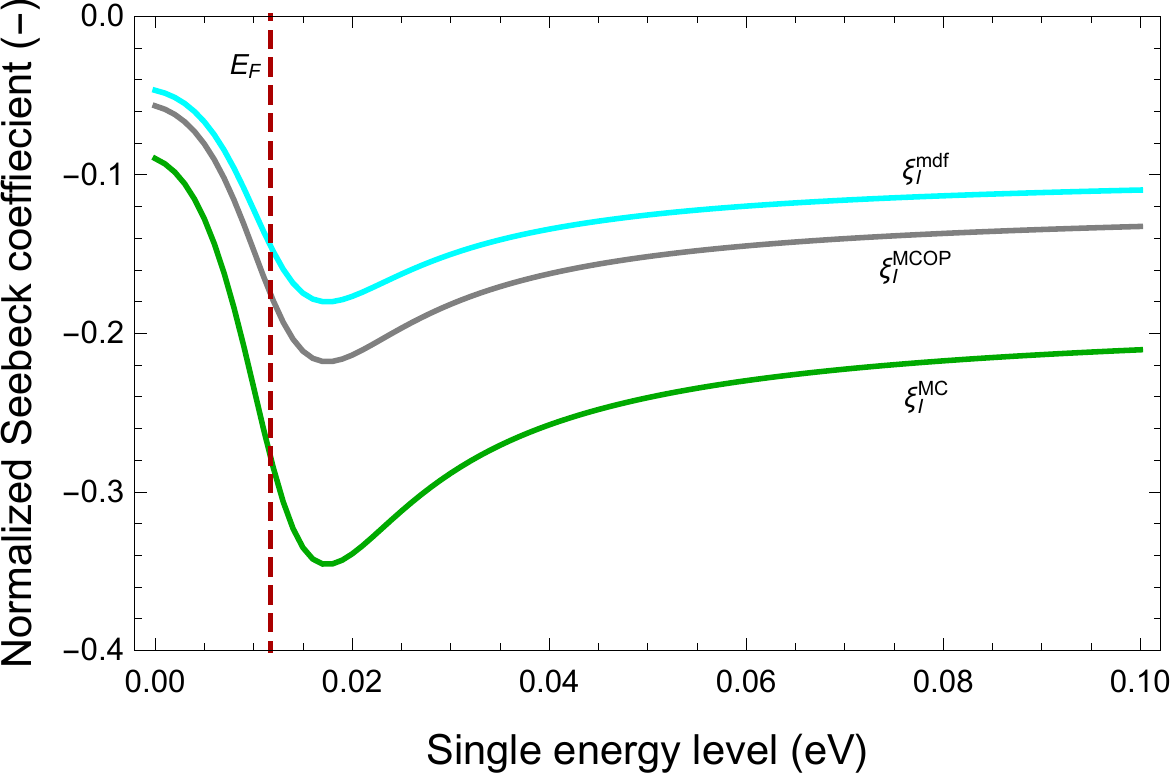}\par}
\caption{Graphs of the normalized SL coefficient family $\nicefrac{\xi_{I}^{\#}}{\left(\frac{k_{B}}{e}\right)}$ evaluated for the two working regimes of an inverse linear energy converter ($mdf$, $MCOP$ and $MC$), plotted as a function of the single resonance energy position in the well $E_{0}$. In a), the case of symmetric barriers is considered, $\Gamma=7\,\textrm{mV}$, $q=0.98$ with temperatures at $20\,\textrm{K}$ and $4\,\textrm{K}$ in descending order. In b), the case of asymmetric barriers is shown, $\Gamma_{L}=7.5\,\textrm{mV}$, $\Gamma_{R}=8.5\,\textrm{mV}$, $q=0.98$ for the same values of both temperatures.}
\label{fig:famseebcoefinv}
\end{figure}

In fig. \ref{fig:famseebcoefinv} 
the curves that represent the behavior of the family of SL coefficients for each operating regime are shown, $\xi_{I}$ represents the measured SL coefficient at $J_{e}=0$. The behavior of this family is analogous to those of the direct linear converter, that is, they have a minimum for one value of the energy $E_{0}$. For any of these values compared to the energy $E_{F}$, the characteristic SL coefficients are negative. All the above tell us that the effective electron current is now produced from the emitter at temperature $T_{R}$ to the collector at temperature $T_{L}$. For a fixed value of Fermi energy ($E_{F}$), a shift in resonance energy is observed (see Fig. \ref{fig:famseebcoefinv}), which means an increase or decrease in the conduction band of any nanostructure. It is also verified that the optimum value of the single resonance energy $E_{0}$ approximates the Fermi energy if $T\rightarrow0$. 

By considering the same heterostructure $Al_{x}Ga_{1-x}As/GaAs/Al_{x}Ga_{1-x}As$ embedded between two reservoirs of $n-GaAs$, and the same value for Fermi energy $E_{F}=11.7\,\textrm{meV}$, we generated another family of Thomson's second relations that reflect the physically reachable operation modes for a refrigerator,
\begin{equation}
\begin{array}{c}
\Pi_{I}^{mfd}=T\xi_{I}\\ \\
\Pi_{I}^{MCOP}=\frac{q^{2}}{1+\sqrt{1-q^{2}}}T\xi_{I}\\ \\
\Pi_{I}^{MC}=\frac{2q^{2}}{4\left(1+\sqrt{1-q^{2}}\right)-q^{2}}T\xi_{I}
\end{array}.\label{eq:optcoolpowe}
\end{equation}

With the introduction of operating regimes as extra-thermodynamic information for a system (linear energy converter), it has been possible to generate a family of second Thomson--like relations \cite{GonzalesArias19}, which are associated with particular steady states. Nevertheless, it is difficult to use this information to pick and choose materials whose heterostructures, in particular RTDs, behave as thermionic generators or as refrigerators.

\section{A new selection criterion for the optimal operation of an RTD as generator or refrigerator}
\label{sec:5}
From a conceptual point of view the direct transport coefficient $\kappa$, can be defined when the flux $J_{e}$ is equal to zero or when the direct force $X_{e}$ is canceled. Based on the experimental fact that thermal conductivity of materials plays an important role in the transport of charge carriers, several authors have proposed and considered a figure of merit ($Tz$), containing the information of the materials \cite{Mahan98,Ioffe57,SnyderToberer08,Bell08}. 

From the two ways to measure thermal conductivity, we have:
\begin{equation}
\kappa_{J}\equiv\left.\mp\frac{J_{Q}}{\nabla T}\right|_{J_{e}=cte}=\frac{\pm L_{11}L_{22}-L_{12}^2\left[\zeta_{D,I}\left(q,x_{D,I}\right)-1 \right]}{T^{2}L_{11}},
\label{eq:thermconducts1}
\end{equation}
and by replacing Eqs. \ref{eq:equiv2} and \ref{eq:coefseeb} in Eq. \ref{eq:thermconducts1}, we obtain
\begin{equation}
\kappa_{J}=\left\{ \begin{array}{c}
\kappa_{V}-TG\xi_{D}^{2}\left[1-\zeta_{D}\left(q,x_{D}\right)\right]\\
-\kappa_{V}+TG\xi_{I}^{2}\left[1-\zeta_{I}\left(q,x_{I}\right)\right]
\end{array}\right.,\label{eq:conectkapp}
\end{equation}
since $\kappa_J > 0$, then $0 < TG\xi_{D}^{2}\left[1-\zeta_{D}\left(q,x_{D}\right)\right] < \kappa_V$ and  $TG\xi_{I}^{2}\left[1-\zeta_{I}\left(q,x_{I}\right)\right] > \kappa_V > 0$. Eq. \ref{eq:conectkapp} can be rewritten as:
\begin{equation}
\kappa_{J}=\kappa_{V}\left[\pm 1\mp \frac{TG\xi_{D,I}^{2}\left[1-\zeta_{D,I}\left(q,x_{D,I}\right)\right]}{\kappa_{V}}\right],\label{eq:newkapp}
\end{equation}
the upper signs correspond to the operation as a generator while the lower ones to a refrigerator. And a kind of figure of merit is obtained,
\begin{equation}
Tz'_{D,I}\equiv Tz'\left(q,x_{D,I}\right)=\frac{TG\xi_{D,I}^{2}}{\kappa_{V}}\left[1-\zeta_{D,I}(q,x_{D,I})\right]=Tz'_{t}\left[1-\zeta_{D,I}(q,x_{D,I})\right].\label{eq:merifig}
\end{equation}

Note that this kind of figure of merit depends on the steady state in which an RTD is operating (see Eqs. \ref{eq:optcoefseebdir} and \ref{eq:optcoefseebinv}). The minimum dissipation function regime leads us to
\begin{equation}
Tz'_{t}=\frac{TG\xi_{D,I}^{2}}{\kappa_{V}},\label{eq:figmeritrad}
\end{equation}
since $\zeta_{D,I}\left(q,x_{D,I}^{mfd}\right)=0$ (see Tab. \ref{tab:optvalforcrat}, and Eqs. \ref{eq:fluxsdir} and \ref{eq:fluxsinv}).

There is a connection between $Tz'_{D,I}$ and the traditional figure of merit $Tz_{D,I}=\nicefrac{TG\xi_{D,I}^2}{\kappa_J}$ \cite{Mahan98}, for the general case we get:
\begin{equation}
Tz_{D,I}=\frac{Tz'_{D,I}}{1\mp Tz'_{D,I}}\left[1-\zeta_{D,I}(q,x_{D,I})\right].\label{eq:equfigmers}
\end{equation}
As a consequence of Eq. \ref{eq:merifig}, this expression includes the operation of an RTD in steady states other than the knowing minimum entropy production steady state, in contrast with the equation reported by G. D. Mahan (see Eq. 1.6 of \cite{Mahan98}). In the operation of energy converters, it has been possible to identify operation zones that reveal good performance during the energy conversion process, as well as moderate dissipation. In what follows, we will present an optimal sub-interval to achieve physical values of $Tz'_{D,I}$ in thermionic generators or refrigerators.

\subsection{RTD as generator}
\label{sec:5.1}
\begin{figure}
\centering
\resizebox{0.7\textwidth}{!}{
\includegraphics[width=7.2cm]{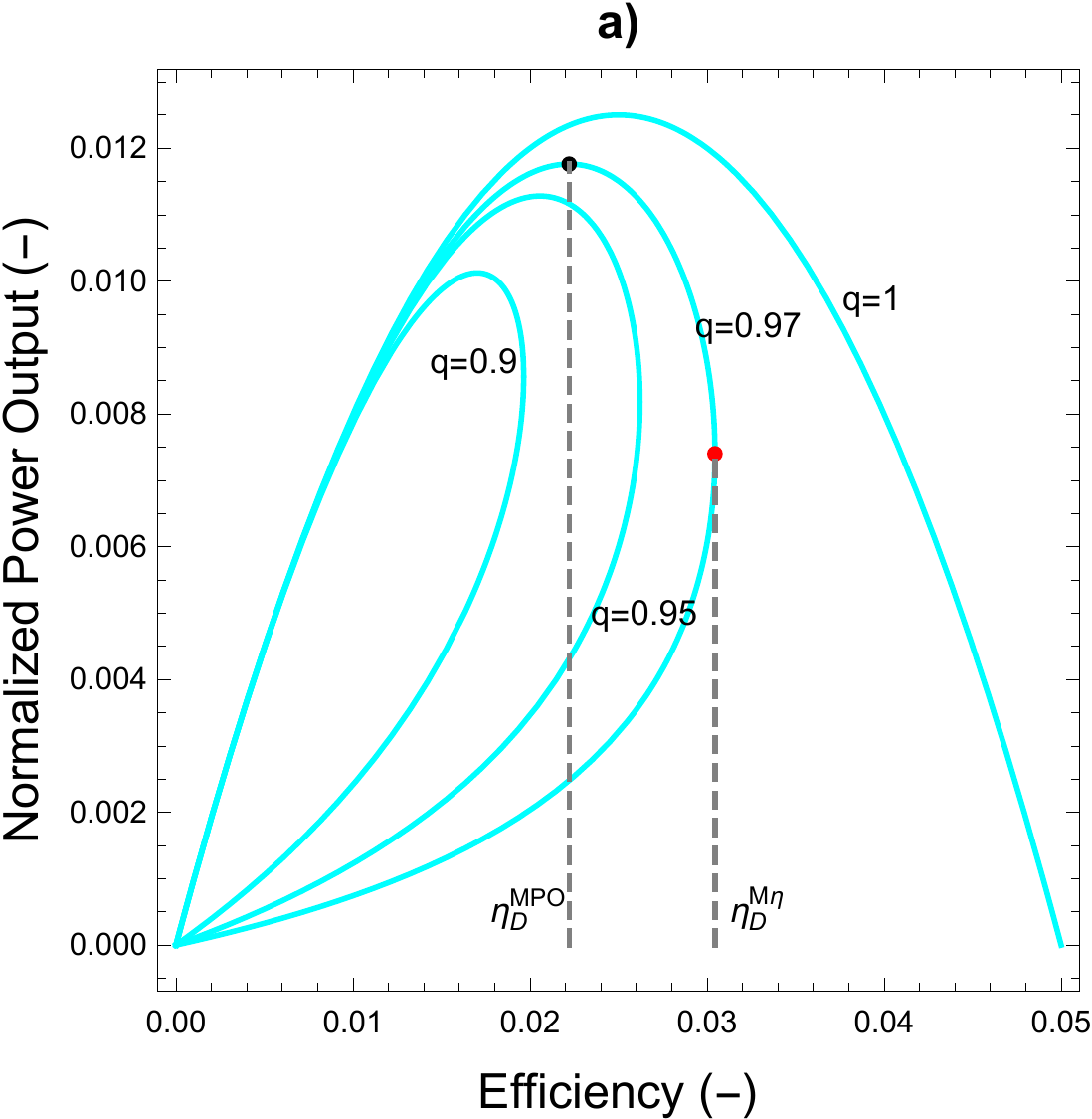}\includegraphics[width=7cm]{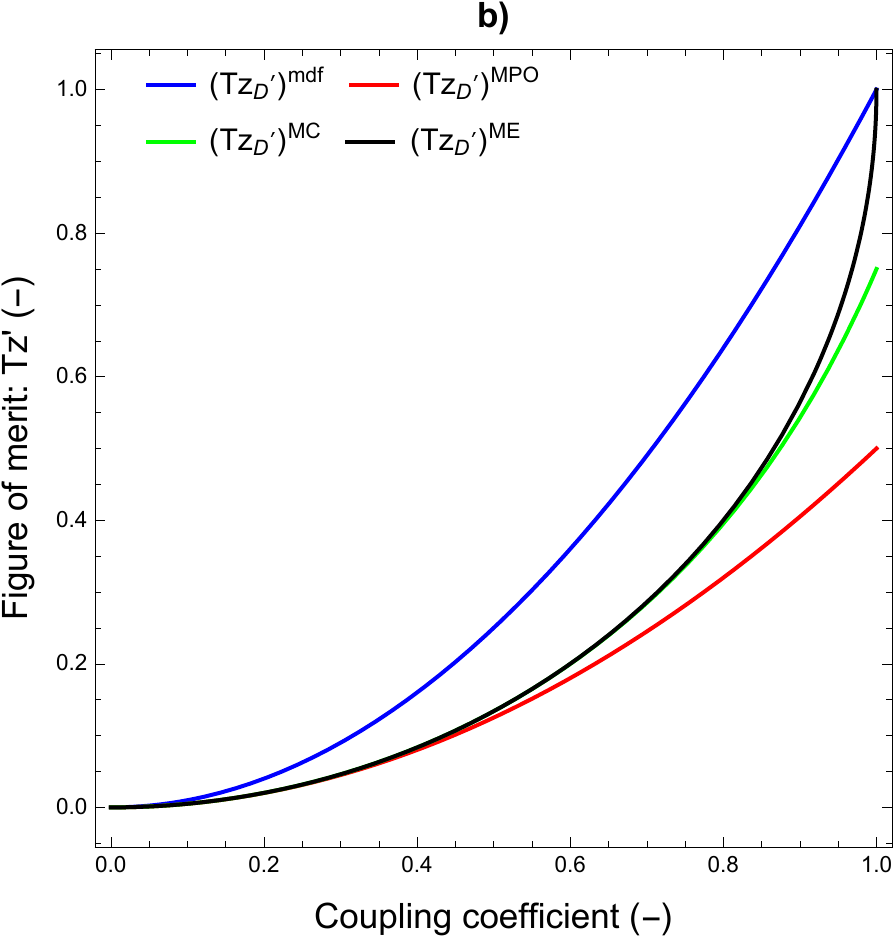}}
\caption{In a), parametric curves $P_{D}^{*}$ vs. $\eta_{D}$ considering different values of $q$'s and $\tau=\nicefrac{T_{R}}{T_{L}}=0.95$ are shown. $\eta_{D}^{MPO}$ and $\eta_{D}^{M\eta}$ express the efficiency values evaluated at the maximum power output and maximum efficiency regimes, respectively. In b), $Tz'_{D}$ is sketched under the optimal operating regimes.}
\label{fig:poetaconvdir}
\end{figure}

From the point of view of energetic optimization, via the described thermodynamic functions in Section 3, we can characterize an optimal zone of operation (see Fig. \ref{fig:poetaconvdir}a). For any linear energy converter that operates directly, we can express the normalized power output in the form $P_{D}^{*}=P_{D}^{*}(\eta_{D},q)$, then we will be able to obtain the parametric curves in the space $P_{D}^{*}$
vs. $\eta_{D}$.

When $x_{D}=\nicefrac{-(\eta_{D}+\eta_{rev})q\pm R_{1}}{2\eta_{rev}}$, the RTD generator power output is expressed as follows:
\begin{equation}
P_{D}^{*}(\eta_{D},q)=\frac{\eta_{D}\left\{ 2\eta_{rev}-q\left[q\left(\eta_{D}+\eta_{rev}\right)\pm R_{1}\right]\right\} }{2\eta_{rev}}L_{22}X_{D2},\label{eq:parampvsefic}
\end{equation}
where $R_{1}=\sqrt{q^{2}(\eta_{D}+\eta_{rev})^{2}-4\eta_{D}\eta_{rev}}$. For each of the loops in Fig. \ref{fig:poetaconvdir}a, we can identify a region bounded by the characteristic functions at $MPO$ and $M\eta$ regimes. This region is characterized by optimal operating efficiency as well as high power output. From Eq. \ref{eq:newkapp} it is stated that in this type of coupled processes $\kappa_{J} < \kappa_{V}$, then $Tz'_{D} < Tz'_t < 1$ \cite{Mahan98}. Therefore, it would guarantee the optimal operation of an RTD as generator (see Fig. \ref{fig:poetaconvdir}b),within the interval
\begin{equation}
Tz'\left(q,x_D\right)\in\left[\frac{q}{2}Tz'_{t},\frac{q}{\left(1+\sqrt{1-q^2}\right)}Tz'_{t}\right],\label{eq:tegbound}
\end{equation}
when $q\rightarrow1$, the system reaches the regime characterized by minimum entropy production. On the other hand, from Eqs. \ref{eq:fluxsdir} and \ref{eq:merifig} we can express $Tz'_{D}$ explicitly in terms of $x_D$ and $q$ as follows:
\begin{equation}
Tz'\left(q,x_D\right)=q^2\left[1-\left(1+\frac{x_D}{q}\right)\right]=-qx_{D},\label{eq:figmerdir}
\end{equation}
where $Tz'_{t}=q^2$ by using Eqs. \ref{eq:equiv2}. In particular, the efficiency of an RTD generator (Eq. \ref{eq:efficcop}) in terms of $Tz'_{D}$ is:
\begin{equation}
\eta_{D}=-\eta_{rev}\frac{x_{D}\left(x_{D}+q\right)}{1+qx_{D}}= \eta_{rev}\left[\frac{Tz'_{D}\left(q^2-Tz'_{D}\right)}{q^2\left(1-Tz'_{D}\right)}\right]
.\label{eq:efficmerfig}
\end{equation}

\subsection{RTD as refrigerator}
\label{sec:5.2}
\begin{figure}
\centering
\resizebox{0.7\textwidth}{!}{
\includegraphics[width=7cm]{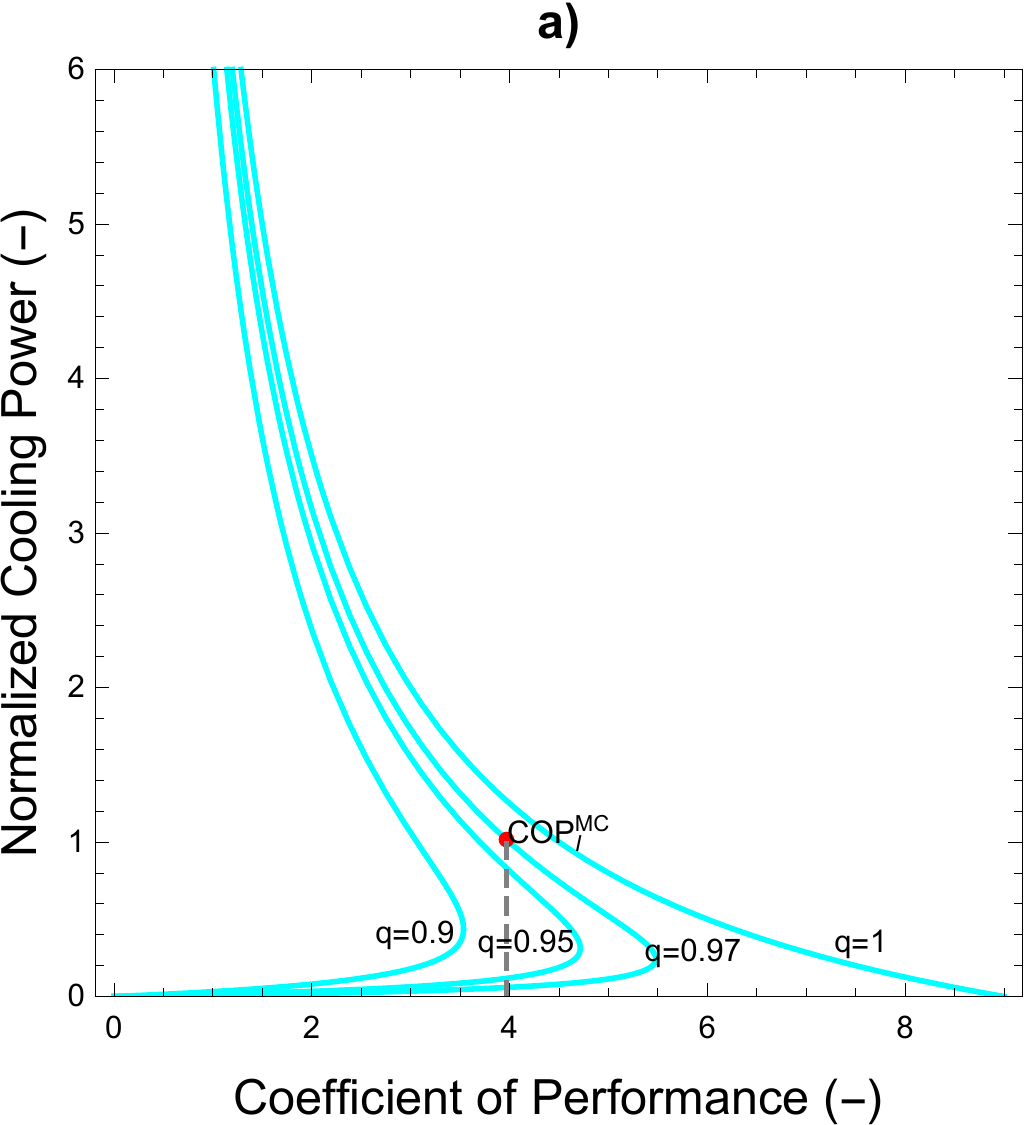}\includegraphics[width=7.3cm]{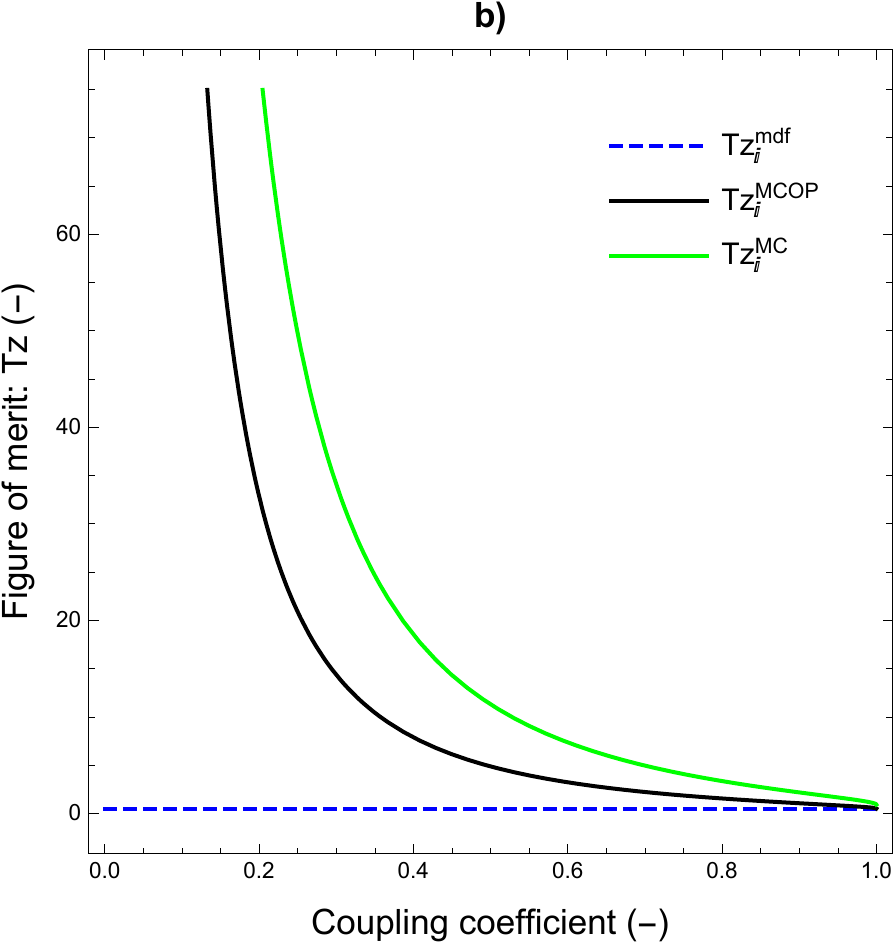}}
\caption{In a), parametric curves $Q_{CI}^{*}$ vs. $COP_{I}$ considering different values of $q$ and a value of $\tau=\nicefrac{T_{R}}{T_{L}}=0.95$. $COP_{I}^{MC}$ expresses the performance coefficient value calculated at the maximum compromise fuunction regime, while $Q_{CI}^{*Max}\rightarrow\infty$. In b), $Tz_{I}$ is sketched under the optimal operating regimes.}
\label{fig:poenfcopconvinv}
\end{figure}

Although there are few thermodynamic functions that provide optimal operating regimes for a linear inverse energy converter, we can characterize enough optimal operating points when we express the normalized cooling power as a function of $COP_{I}$, i. e., we can build the parametric curves in the space $Q_{CI}^{*}$ vs. $COP_{I}$ (see Fig. \ref{fig:poenfcopconvinv}a).

In this case, if $x_{I}=\nicefrac{-(COP_{I}-COP_{rev})\pm R_{2}}{2COP_{I}}$, the cooling power of a thermionic refrigerator is written as:
\begin{equation}
Q_{CI}^{*}(COP_{I},q)=-\frac{\left(COP_{I}-COP_{rev}\right)q\pm R_{2}}{\left(COP_{I}+COP_{rev}\right)q\pm R_{2}},\label{eq:paramqcvscop}
\end{equation}
where $R_{2}=\sqrt{q^{2}(COP_{I}+COP_{rev})^{2}-4COP_{I}COP_{rev}}$. The curves shown in Fig. \ref{fig:poenfcopconvinv}a have
two regions separated by the $MCOP$ regime. Values for cooling power greater than $Q_{CI}^{MCOP}$ are in an optimal region, which has high performance coefficient values and moderate cooling power. In this case, the coupled processes lead to $\kappa_{J} > \kappa_{V}$, then $Tz'_{I} > Tz'_t$ but the optimal operating interval for a RTD as refrigerator is better displayed with the traditional figure of merit (see Fig. \ref{fig:poenfcopconvinv}b):
\begin{equation}
Tz\left(q,x_I\right)\in\left[Tz_{I}^{mdf},Tz_{I}^{MC}\right],\label{eq:tecbound}
\end{equation}
since $Tz'\left(q,x_I\right)=\nicefrac{q}{x_{I}}$. Therefore, it is also possible to express $Tz_{I}$ in terms of $x_I$ and $q$ (see Eqs. \ref{eq:fluxsinv} and \ref{eq:equfigmers}),
\begin{equation}
Tz\left(q,x_I\right)=\frac{-q}{\left(x_{I}-q\right)}\left[1-\left(1+\frac{1}{qx_I}\right)\right]=\frac{1}{x_I\left(x_{I}-q\right)}.\label{eq:figmerinv}
\end{equation}
In this case, the operating coefficient of an RTD refrigerator (Eq. \ref{eq:efficcop}) now in terms of $Tz'_{I}$ is:
\begin{equation}
COP_{I}=-COP_{rev}\frac{x_{I}\left(x_{I}+q\right)}{1+qx_{I}}=
COP_{rev}\left[\frac{q^2\left(1-Tz'_{I}\right)}{q^2-Tz'_{I}}\right]
.\label{eq:copmerfig}
\end{equation}

\section{Conclusions}
\label{sec:6}
Despite the advances in the construction and characterization of various microelectronic devices, the processes of charge carrier transport that occur inside them are associated with the inevitable heat transfer. Within the context of Linear Irreversible Thermodynamics, any system whose energy conversion processes are described by transport equations, is a good candidate to study its energy performance under the scheme of linear energy converters. Resonant Tunnel Diodes (RTDs), put in the context of irreversible processes, can be seen as thermionic devices with coupled processes. Its application is not restricted exclusively to the information transmission, according to its objective they can also work as electric current generators or as refrigerators.

In this manuscript, we have verified, the Landauer--Büttiker (LB) approach that describes the transport of charge carriers and energy into an RTD. By considering elastic scattering, they can be studied in the linear regime and the phenomenological equations of Onsager fullfilled. When the extra--thermodynamic information associated with those systems is taken into account,  a family of Thomson--like relation that do not necessarily correspond to the open circuit condition ($J_{D1,2}(x_{D,I}^{mfd},q)=0$) is revealed. In an RTD that is formed by a heterostructure of $Al_{x}Ga_{1-x}As$, we observe that the resonance level, in this type of single resonance energy selective electron system, approximates the Fermi energy $E_{F}=11.7\,\textrm{meV}$ when the reference temperature of the system is very small. In addition, by taking different values for the leakage of electrons between the barriers $\Gamma_{L}\neq\Gamma_{R}$, the resonance energy $E_{0}^{max}$ decreases. The information of the operating regimes can also be reflected in the behavior of the SL coefficient, i. e., we found a hierarchy between these regimes, for an RTD generator: $\xi_{D}^{MPO}<\xi_{D}^{MC}<\xi_{D}^{M\eta}<\xi_{D}^{mdf}$ while for an RTD refrigerator $\xi_{I}^{MC}<\xi_{I}^{MCOP}<\xi_{I}^{mdf}$. Both forms of characterizing the SL coefficients have PL coefficients ($\Pi$) associated, which also reflects the information of the operating modes.

With the purpose of finding a new selection criterion for optimal operation of an RTD, we used the definitions of the thermal conductivities $\kappa_V$ and $\kappa_J$, through a kind of figure of merit: $Tz'_{D,I}$, which was partially identified by Mahan \cite{Mahan98} by using another analysis. This new figure of merit has an advantage over the obtained in \cite{Mahan98} and the traditional one ($Tz_{mfd}$). This new criterion contains the operation modes physically achievable by any energy converter. Additionaly, we found a functional relationship for the two generalized versions of these figures of merit (Eq. \ref{eq:equfigmers}). In order to extend the utility of the operation modes, we have found the operating bounds to ensure that an RTD operates as a current generator or as a refrigerator. These bounds are related to the operation of single resonant electron devices and could help to reduce the search for semiconductor materials to cut down its growth costs via $Tz'_{D,I}$.

\section*{Acknowledgement}

Thanks to F. Angulo-Brown for his recommendations to improve the manuscript and M. Moreno for stimulating discussion.

This work was partially supported by CONACYT Grants: 288669 
Instituto Polit\'ecnico Nacional: SIP-project number: 20195925,
COFAA-Grant: 5406, EDI-Grant: 1750 and SNI-CONACYT Grant: 16051, MEXICO.

\end{document}